# Imaging Self-aligned Moiré Crystals and Quasicrystals in Magic-angle Bilayer Graphene on hBN Heterostructures

Xinyuan Lai,[1] Daniele Guerci,[2] Guohong Li,[1] Kenji Watanabe,[3] Takashi Taniguchi,[3] Justin Wilson,[4] Jedediah H. Pixley,[1,2,5] Eva Y. Andrei[1*]

[1]Department of Physics and Astronomy, Rutgers University, 136 Frelinghuysen Road, Piscataway, NJ 08854, USA
[2]Center for Computational Quantum Physics, Flatiron Institute, 162 Fifth Avenue, New York, NY 10010, USA
[3]Research Center for Functional Materials, National Institute for Materials Science, 1-1 Namiki, Tsukuba 305-0044, Japan
[4]Department of Physics and Astronomy, and Center for Computation and Technology, Louisiana State University, Baton Rouge, LA 70803, USA
[5]Center for Materials Theory, Rutgers University, 136 Frelinghuysen Road, Piscataway, NJ 08854, USA

∗To whom correspondence should be addressed, E-mail: eandrei@physics.rutgers.edu

**Abstract:** Using scanning-tunneling-microscopy and theoretical modeling on heterostructures of twisted bilayer graphene and hexagonal Boron-Nitride, we show that the emergent super-moiré structures display a rich landscape of moiré-crystals and quasicrystals. We reveal a phase-diagram comprised of commensurate moiré-crystals embedded in swaths of moiré quasicrystals. The 1:1 commensurate crystal, expected to be a Chern insulator, should only exist at one point on the phase-diagram, implying that it ought to be practically undetectable. Surprisingly we find that the commensurate crystals exist over a much wider than predicted range, providing evidence of an unexpected self-alignment mechanism that is explained using an elastic-network model. The remainder of the phase-diagram, where we observe tunable quasicrystals, affords a new platform for exploring the unique electronic-properties of these rarely found in nature structures.

**One-Sentence Summary:** Interfering moiré patterns in twisted structures produces Matrioshka-doll geometries that are otherwise not possible in nature.

**Main text:**

Twisting a pair of atomically thin layers produces a tunable moiré pattern with profound implications for the electronic properties (*1-5*). By twisting three or more layers, these moiré patterns can interfere, generating novel geometric arrangements unavailable with two layers. In GG, a twist of $\theta_{GG} \approx 1°$ between the layers - known as magic angle - produces a nearly flat band with non-trivial topology (*6*). Breaking either the time-reversal symmetry with a magnetic field or the sublattice symmetry ($C_{2z}$) with a staggered potential produces Chern insulating states at integer fillings of the moiré unit cell(*7-11*). However, in contrast to the readily observed Chern insulators in an external magnetic field (*10*), those induced by breaking $C_{2z}$ symmetry are rare, their hallmark Anomalous Quantum Hall effect (AQHE) having been reported in only two samples (*7, 9*) that were encapsulated in hexagonal boron nitride (hBN) with one of the graphene layers aligned to one of the hBN crystals while the other remained unaligned. The Boron and Nitrogen atoms occupy different sublattices of the hexagonal structure, so that aligning the two crystals breaks the $C_{2z}$ symmetry (*12-14*) of graphene. Because of the lattice mismatch ($a_G = 0.2461 nm$ and $a_{BN} = 0.2504 nm$ for graphene and hBN, respectively), alignment between magic angle GG and hBN can only be achieved for three twist angles: $\theta_{GBN}^0 = \pm 0.55°, 0°$ (*15-18*). While techniques exist for precisely aligning two graphene layers cut from the same flake(*19*), aligning graphene and hBN over a several-micron sample area is not yet possible. This raises the question of what enabled the rare observation of the AQHE in magic angle GG aligned with hBN.

Using low-temperature scanning-tunneling-microscopy (STM) (*20, 21*), we study moiré structures in GG/hBN devices comprised of twisted bilayer graphene deposited on a bulk hBN ( ~30nm) substrate, with $\theta_{GG} \approx 1°$ and $\theta_{GBN} < 1°$ (Fig. 1A). These devices notably and necessarily lack the randomly aligned top layer of hBN used in the AQHE transport experiments. We find a strong tendency of the lattices to self-align into the preferred AAB stacking order, where the AA sites of the GG moiré lattice are aligned with only Boron atoms in the hBN. Surprisingly, the aligned lattices are observed over a broader range of twist angle pairs than expected for rigid lattices, suggesting that the effects of broken $C_{2z}$ symmetry, including the AQHE, should be readily observed. We describe this lattice relaxation theoretically by constructing a classical Frenkel-Kontorova model (FK-model) (*22*) which treats the AA moiré sites of GG as a classical elastic network whose relaxation on the potential produced by the twisted hBN substrate produces large commensurate domains on the moiré-of-moiré scale. We discover a moiré phase diagram that describes the structural arrangement of GG comprised of lines of commensurate moiré lattices embedded in a background of quasicrystals(*23-30*). Last, we show that the competition between the elastic energy cost of alignment and the van der Waals energy gain of the preferred AAB stacking controls the emergence of self-aligned domains.

**Preferred GG-GBN stacking order**

Commensurate moiré lattices are characterized by one of three distinct periodic local stackings between the AA sites of GG and the underlying hBN lattice(*15-18*). These are labeled AAB, AAN, or AAC corresponding to AA aligned with either the B atom, the N atom or the center of the hBN unit cell, as illustrated in Fig. 1B, 1C, and 1D, respectively. The stacking pattern changes by shifting the hBN relative to the graphene layer adjacent to it. This is in stark contrast to the moiré pattern in GG which is insensitive to a shift between the layers.

The stacking order is revealed in STM topography by imaging the two sets of moiré patterns simultaneously, as illustrated in Fig. 1. Our study, comprised of dozens of twist angle pairs, reveals large areas of contiguous AAB stacking, indicating commensurability, even for samples whose twist angles don't match the rigid lattice conditions, $\theta_{GG}^0 = 1.1°; \theta_{GBN}^0 = \pm 0.55°$ for which both moiré lattices have the same wavelength $L_M = 12.8\ nm$. Despite being quite far from the 1:1 rigid lattice commensuration condition, we observe well-aligned AAB domains that extend over many moiré cells. This suggests a relaxation mechanism that causes the GG and GBN moiré patterns to self-align. Indeed, according to first principle calculations(*31, 32*), the AA sites of GG have the highest onsite energy while the $C_B$ sites of GBN have the lowest energy, and the AAB configuration is likely to be energetically favored. This assumption is supported by comparing the topography obtained from forward and backward STM scans in regions with different stacking orders (Fig. S2). The topography of AAB domains is identical in forward and backward scans, while it changes with scan direction for other stacking orders. A relative shift of the local stacking triggered by the interaction between the sample and moving STM tip for non-AAB stacked sites indicates instability. Observing large AAB-ordered domains versus disordered regions with non-AAB stacked unit cells further validates this conclusion.

**Moiré commensurate-incommensurate transition**

All data presented are within the small angle regime such that the two moiré wavelengths are given by: $L_{GG} = \frac{a_G}{2sin\left(\frac{\theta_{GG}}{2}\right)} \sim \frac{a_G}{\theta_{GG}}$ and $L_{GBN} = \frac{(1+\delta)a_G}{\sqrt{2(1+\delta)(1-\cos(\theta_{GBN}))+\delta^2}} \sim \frac{a_{BN}}{\sqrt{(\theta_{GBN})^2+\delta^2}}$, where $\delta = \frac{a_{BN}}{a_G} - 1 = 0.017$ is the lattice mismatch. Thus, the maximum moiré wavelength for perfectly aligned lattices is

14.7 nm for GBN and unbounded for GG. Experimentally, we obtain the moiré lattice constants from topography images and their Fourier transforms (FFT). From the FFT we calculate the moiré reciprocal lattice vectors, $|\vec{K}_{GG}| = 2/(\sqrt{3}L_{GG})$ and $|\vec{K}_{GBN}| = 2/(\sqrt{3}L_{GBN})$ by averaging over the respective sets of six Bragg peaks, from which we extract the average moiré wavelengths. We studied samples within the range $L_{GG} > 3.5\ nm$ and $L_{GBN} > 10.2\ nm$.

Patterns where the two moiré lattice vectors coincide, corresponding to 1:1 commensurate alignment, share a common moiré wavelength which we label as $L_M$. In Fig. 2A and Fig. 2C, we illustrate two commensurate patterns with $L_M = 11.1\ nm$ and $12.8\ nm$, respectively. In both cases, the topography reveals a triangular lattice of AAB stacking points spanned by two primitive lattice vectors (white/blue arrows). The FFTs of these structures (Fig. 2B, Fig. 2D) display a set of 6 primary Bragg peaks produced by the perfectly aligned GG and GBN moiré patterns which are accompanied by 12 second-order peaks. Additional examples of 1:1 alignment are shown in Fig. S3.

In addition to the self-aligned moiré lattices, we observe incommensurate lattices – Fig. 2E and Fig. 2G - that we classify as quasicrystals: they exhibit well-defined Bragg peaks and are spanned by more than two independent basis vectors in contrast to only two for 2D Bravais lattices (*33, 34*). Describing the moiré quasicrystals observed here requires four independent basis vectors, one pair spanning the GG lattice and the other the GBN lattice. In the example shown in Fig 2E, the pattern is characterized by two independent pairs of basis vectors (white and blue arrows). From the two corresponding sets of Bragg peaks (white and blue circles in Fig. 2F), we obtain $L_{GG} = 7.1\ nm$; $L_{GBN} = 10.9 nm$. We note that the electrons experience both crystal lattices leading to interference effects which produce additional peaks beyond the Bragg peaks expected for non-interacting crystals. This becomes apparent in energy-resolved measurements that will be discussed elsewhere.

For moiré wavelengths pairs that approach each other and $L_{GG} < L_{GBN}$, we observe a moiré-of-moiré pattern with a super-period labeled $L_{MM}$(*35*). This pattern is visible in Fig. 2G whose FFT in Fig 2H shows two sets of peaks corresponding to the slightly mismatched GG and GBN moiré lattice vectors, but the super-period Bragg peaks are not resolved. An extreme case where the two moiré lattice vectors are close but not equal is shown in Fig. 2I and Fig. 2H. Here, the GG peaks are smeared out into six diffuse clouds indicating structural frustration and the loss of symmetry that is evident in the topography and the associated Bragg peaks. More examples of incommensurate patterns are shown in Fig. S4. We classify a pattern as a quasicrystal by using the absence of a commensurate length scale, $L_c \equiv nL_{GG} + mL_{GBN}$ where *n, m* are integers, within the sample boundaries. To this end we calculate the ratio $\rho = L_{GG} / L_{GBN}$ and take the closest rational number $\rho \approx n/m$ to define the shortest commensuration length scale $L_c$. Patterns with $L_c$ larger than the sample size are classified as quasicrystals. We note that there is a separation of length scales with $L_c$ being at least one order of magnitude larger than $L_{GG}\ or\ L_{GBN}$ for all patterns designated as quasicrystals (Table S2). Examples of quasicrystals are shown and Fig. S4 and analyzed in Table S2. An extreme case where the two moiré lattice vectors are close but not equal is shown in Fig. 2I and Fig. 2H. Here, the GG peaks are smeared out into six diffuse clouds indicating structural frustration and loss of symmetry that is evident in the topography and the associated Bragg peaks.

The collected data is classified in the [$L_{GBN}$, $L_{GG}$] phase diagram shown in Fig. 3A. Commensurate moiré lattices, indicated by black dashed lines, each contain a green star marking the rigid lattice commensuration point which we label as $L_{M0}$. For the 1:1 commensuration, we have $L_{M0} \sim 12.8$nm.

Remarkably, many 1:1 commensurate data points (solid black dots) deviate by as much as 20% from $L_{M0}$, providing clear evidence of self-alignment. Other commensurate double moiré crystals are observed on the 2:1, 1:2, 1:3, 1:4, and 1:5 lines (Extended Data Fig. 5). The data outside of the commensuration lines reveal a plethora of quasicrystals (green dots) formed when the two moiré patterns are incommensurate.

To analyze the data, we assume that only the two graphene monolayers relax, whereas the bulk hBN is rigid. In Fig. 3B, we show the GG/GBN Brillouin zone corresponding to the 1:1 rigid lattice commensuration condition at $\varphi = 120°$, where $|\theta_{GBN}| = \frac{\theta_{GG}}{2}$. In general, misaligned lattices will have $|\theta_{GBN}| \neq \frac{\theta_{GG}}{2}$ as well as $\varphi \neq 120°$. Global commensuration would thus require rescaling of the lattice vectors so that $K_{GG} = K_{GBN} \equiv K_M$, and their relative angle is $120°$ (or $60°$) (*15, 36*). For the symmetric alignment case, where $\varphi = 120°$ and $|\theta_{GBN}| = \frac{\theta_{GG}}{2}$ (Fig. 3C, 3D top panels) both graphene layers stretch or contract by the same amount corresponding to a reciprocal lattice vector change for both top and bottom graphene layers of $\Delta K_G = \frac{\sqrt{3}}{2}\left(K_{GG0} - K_{GG}\right)$ and to a symmetric atomic strain $\varepsilon_s = \frac{\Delta K_G}{K_{G0}} = \frac{\sqrt{3}}{2}\frac{a_G}{L_M}\left(\frac{\Delta L_M}{L_{M0}}\right)$. For $|\theta_{GBN}| \neq \frac{\theta_{GG}}{2}$ (Fig. 3C bottom panel) the alignment condition is reached through additional amounts of strain added to $\varepsilon_s$ with opposite signs in the two layers, resulting in the asymmetric strain $\varepsilon_a$, as illustrated in Fig 3D where $\varepsilon_a$ in the top layer is three times larger than in the bottom layer (*36*). The elastic energy cost per moiré cell for aligning the two lattices, $E_{el}$, as a function of the twist angle ratio for the 120° case (*36*), is plotted in Fig. 3E and 3F, and $E_{el}(L_M)$ for both 60° and 120° cases is plotted in Fig. 3G. Projecting the measured $L_M$ values of the 1:1 aligned lattices (black squares) onto the calculated curves, we obtain the range of $E_{el}$ for which 1:1 alignment is observed. The upper bound of $E_{el}$ is indicated by the pink line in Fig. 3G. Interestingly, the 120° alignment is energetically favorable for most experimentally observed commensurate patterns, except when they coincide with the rigid lattice commensuration condition.

### Self-alignment induced moiré superstructures.

The global self-alignment model discussed above was based on several assumptions: homogeneous relaxation, a uniform twist angle, and no externally imposed strain. In experimental samples however, slight deviations from these conditions can disrupt the alignment by creating non-aligned boundary regions that separate aligned domains, as illustrated in Fig. 4A. These boundaries can be classified as tensile (dashed black line) or shear type (dashed green line), depending on their orientation relative to the crystallographic directions. In the zoomed-in image of the two types of domain boundaries, Fig. 4B and 4C, the GBN unit cells are outlined by dashed hexagons, and black spots mark the AA sites. At the domain boundaries, the hexagons representing moiré unit cells are distorted, and the AA sites are no longer aligned to the center of the dark $C_B$ sites as for the equilibrium AAB stacking. For the tensile boundary (Fig. 4B), the displacement of the bright AA spots relative to dark spot centers is perpendicular to the boundary line creating a row of "empty" GBN unit cells, while for a shear boundary (Fig. 4C), the displacement is parallel to the boundary. The width of the domain wall is determined by the competition between the intralayer elastic energy and the interlayer adhesion energy. The sharp domain wall on the order of few moiré unit cells here suggests a strong tendency for self-alignment.

Earlier reports on monolayer graphene supported on hBN(*37*), have shown a commensurate to incommensurate transition similarly driven by the competition between elastic and vdW stacking energies. However, unlike GBN moiré patterns where the graphene and hBN lattice constants are fixed, a crucial difference is that the GG-GBN moiré lattice constants are readily tunable by changing the twist angles. This makes it possible to produce a plethora of new moiré lattices that are inaccessible in the bilayer GBN system.

To study how large aligned domains can form despite their initial incommensuration, we construct a 2D classical FK-model (22) at the moiré scale of GG. The model describes the AA sites of GG as a triangular elastic network with energy $E_{el}[\{\vec{u}\}] = Y \sum_{<i,j>} \frac{(\vec{u}_i - \vec{u}_j)^2}{2}$ where $Y$ is the elastic constant and $\vec{u}_i$, the displacement with respect to the rigid lattice, can relax in the presence of the incommensurate GBN potential $E_{pot}[\vec{r}] = -U_0 \sum_{n=1}^{3} \cos(\vec{K}_{GBN_n} \cdot \vec{r} + \varphi)$. In our simulation, we fix $\theta_{GG} = 1.1°$ ($L_{GG} = 12.8 nm$) in the magic-angle region while varying $\theta_{GBN}$. Over a range of angles, $\theta_{GBN} \in [0.33°, 0.78°]$, we find that the relaxed lattice structure shows commensurate domains whose characteristic size which is of order the moiré-of-moiré wavevector $L_{MM} \sim 1/|K_{GG} - K_{GBN}|$, decreases with the deviation from the rigid lattice commensuration condition $\theta_{GBN} \sim \pm\, 0.55°$. The evolution of the commensurate domain size with $\theta_{GBN}$ is illustrated in Fig S11. As expected, the domain size is unbounded at the rigid lattice commensurability condition $\theta_{GBN} \sim 0.55°$ (Fig. 4E) and it decreases for both smaller and larger values of $\theta_{GBN}$, where either $L_{GG} \lesssim L_{GBN} = 13.2 nm$, ($\theta_{GBN} = 0.48°$ Fig. 4D) or $L_{GG} \gtrsim L_{GBN} = 12.3 nm$ ($\theta_{GBN} = 0.58°$ Fig. 4F). The tensile and shear type domain boundaries observed in the experiment (Fig. 4A) also emerge in the simulation as illustrated in Fig 4D. Outside the interval $\theta_{GBN} \in [0.33°, 0.78°]$ ($L_{GBN} \in [11.5, 13.9]$ nm) there is no separation between the moiré $L_M$ and the super-moiré $L_{MM}$ length scales and relaxation can no longer restore commensuration. To demonstrate this, we first consider two incommensurate rigid (unrelaxed) lattices with four distinct Bravais vectors representing GG and GBN at $\theta_{GBN} = 0.9°$ ($L_{GBN} = 10.3$ nm), see Fig.4E. Allowing these lattices to relax we find a moiré quasicrystal (38) as clearly indicated by the presence of peaks at the expected four non-commensurate reciprocal lattice vectors in the structure factor (Fig. 4I). This theoretical description is in good agreement with the experimental results providing a simple mechanism for explaining the formation and destruction of commensurate domains.

It is useful to highlight again the difference here between the previous findings of the AQHE in two other samples that both have an additional unaligned hBN layer and our STM experiments with only one hBN layer. To model the encapsulated scenario, we include an additional unaligned hBN potential in our simulations. This additional potential frustrates the relaxation mechanism and greatly reduces the regime of stability of moiré-of-moiré commensurate domains. Instead, it enlarges the regime of quasicrystal formation (see Supplemental). We posit that this could help explain the scarcity of encapsulated samples which exhibit AQHE, while the STM experiments reveal a proliferation of self-aligned regions, which could give rise to an AQHE.

In summary, we discovered a rich variety of emergent moiré superlattices in GG-GBN heterostructures and showed that they can be classified in a phase diagram comprising lines of commensurate moiré crystals embedded in a sea of aperiodic lattices including moiré quasicrystals. The 1:1 commensurate moiré crystal, which are expected to be Chern-insulators, should only exist at one point on this phase-diagram, rendering them practically undetectable. Surprisingly, we find that they extend far beyond the commensuration-point, providing direct evidence for a self-alignment mechanism that we expect is responsible for the observation of the AQHE in this system. The self-alignment is driven by the competition between the vdW energy gained in the favorable AAB stacking configuration and the elastic energy cost of this configuration. Our study reveals self-alignment as a previously neglected but fundamental property of twisted multilayer systems. This mechanism allows moiré superlattices to overcome finite incommensurability from layer-by-layer device fabrication by forming commensurate domains whose electronic properties can be described within the Bloch framework of periodic lattices. Self-alignment reduces the sensitivity to twist-angle disorder created during device fabrication and vastly increases the chances of achieving samples with desired global properties(38). The trade-off is the formation of domain boundaries whose presence, which can only be detected with local probes(39, 40), may introduce noise and interfere with the detection of the global properties. Outside of the lines containing the self-aligned moiré crystals, we observe a host of 2D quasicrystal lattices whose description requires more than two independent basis vectors. 2D quasicrystals are rarely found in nature, and their expected correlated phases and topological properties are largely unverified. Our findings provide a clear cut platform for creating and tuning 2D quasicrystals and for exploring their fascinating electronic and many-body properties.

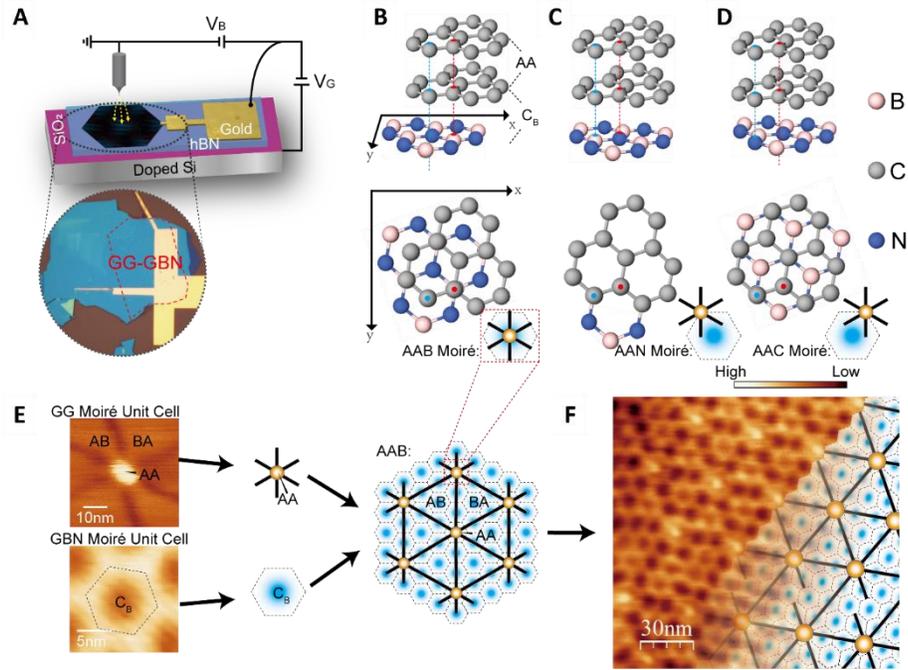

Fig. 1. **Imaging GG and GBN double moiré pattern and the preferred local stacking order.** (**A**) Schematic drawing of the experimental setup. Sample bias $V_B$ is applied to the gold electrode. Sample doping is tunable by $V_G$ applied to the silicon substrate and the moiré pattern is imaged with the scanning tip. Inside the circle is an optical image of the device. (**B-D**), Schematic atomic configuration of different stacking orders: (**B**) AAB, (**C**) AAN, (**D**) AAC. The top and bottom panels correspond to 3D and top views of the model respectively. Boron, carbon, and nitrogen atoms are labeled by pink, grey and blue circles respectively. Red and blue spots mark the graphene A and B sublattices respectively. The lower right insets schematically represent the different sackings as viewed in STM topography and shown in Fig 1E. (**E**) STM topography and schematic drawing of GG and GBN moiré patterns. The GG moiré unit cell shown in the top left panel and schematically to its right, consists of a bright central AA region surrounded by six alternating less bright Bernal stacked (AB/BA) regions. The smaller area of AA compared to AB/BA regions reflects the lattice relaxation towards the energetically favored Bernal stacking (*41, 42*). The bottom left panel and the schematic inset on the right show the STM topography of a single GBN moiré cell consisting of a central dark spot corresponding to the energetically favored $C_B$ stacking (C atoms of one of the graphene sublattices is above B atoms of hBN) whose onsite energy is lower by ~ 10 meV/atom (*31*) compared to the surrounding bright ring. Superposing the two patterns (right panel) schematically shows the expected topography for $L_{GG} = 3L_{GBN}$. Here GG unit cells are represented by a triangular array of bright dots connected by dark sticks, and GBN unit cells are hexagons with $C_B$ sites marked in blue. (**F**) STM topography showing the double moiré pattern produced in an AAB stacking configuration under tunneling conditions where the GBN and GG moiré patterns coexist. The GG and GBN moiré patterns appear as bright and dark spots respectively. Thus, our analysis uses a bright spot overlapping a dark spot as an indicator of local alignment between the two moiré lattices. Here $L_{GG} = 35\ nm$; $L_{GBN} = 11.2\ nm$ ($\theta_{GG} = 0.4°\ and\ \theta_{GBN} = 0.8°$). Sample bias $V_B$ = -500mV, gate voltage $V_G$= 0V, tunneling current I = 20pA.

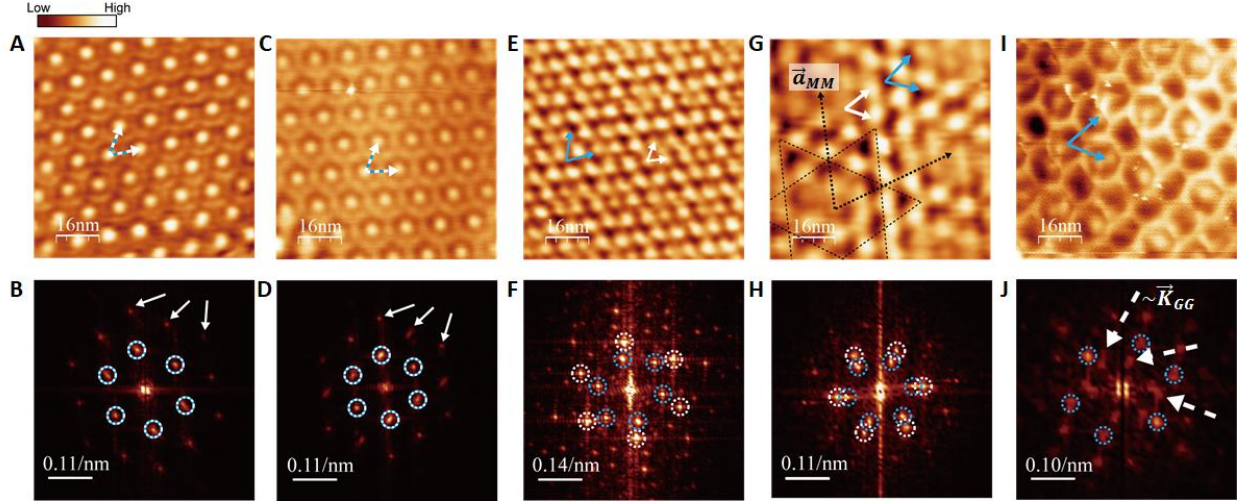

Fig. 2. **Double moiré patterns across the commensurate-incommensurate transition.** STM topography and corresponding FFT of representative GG/hBN double moiré patterns. **(A)** Commensurate pattern: $L_M = 11.1\ nm$ ($L_{GG} = L_{GBN}$). ($\theta_{GG} = 1.27°; \theta_{GBN} = 0.83°$) **(C)** Commensurate pattern: $L_M = 12.8\ nm$ ($\theta_{GG} = 1.10°; \theta_{GBN} = 0.55°$). **(E)** Incommensurate: $L_{GG} = 3.6 nm; L_{GBN} = 11.0\ nm$. ($\theta_{GG} = 3.93°; \theta_{GBN} = 0.86°$) Both GG and GBN periods are visible. **(G)** Incommensurate with visible moiré-of-moiré period: $L_{GG} = 10.7\ nm; L_{GBN} = 14.5\ nm$. ($\theta_{GG} = 1.32°; \theta_{GBN} \approx 0°$). **(I)** Incommensurate with broken translational symmetry: $L_{GG}$ is non-uniform ranging from 12 to 25 nm; $L_{GBN} \approx 12.6\ nm$. The GG and GBN moiré lattice vectors ($\vec{a}_{GG}$ and $\vec{a}_{GBN}$) are marked in **(A-I)** by white and blue arrows. The moiré-of-moiré lattice vectors $\vec{a}_{MM}$ is marked in **(G)** with black arrows. **(B), (D), (F), (H),** and **(J)** are 2D Fourier transforms of **(A), (C), (E), (G),** and **(I),** respectively. Wave vectors corresponding to the Bragg peaks of the GG and GBN moiré patterns are outlined by white and blue circles, respectively. The incommensurate patterns attributed to quasicrystals are discussed in the text. White arrows in **(B)** and **(D)** point to second-order Bragg peaks, and white arrows in **(J)** point to the broadened GG Bragg peaks. Tunneling parameters: $V_B$ = -500mV; $V_G$=0V; I = 20pA. Twist angle values are obtained from moiré wavelengths using the rigid lattice formula.

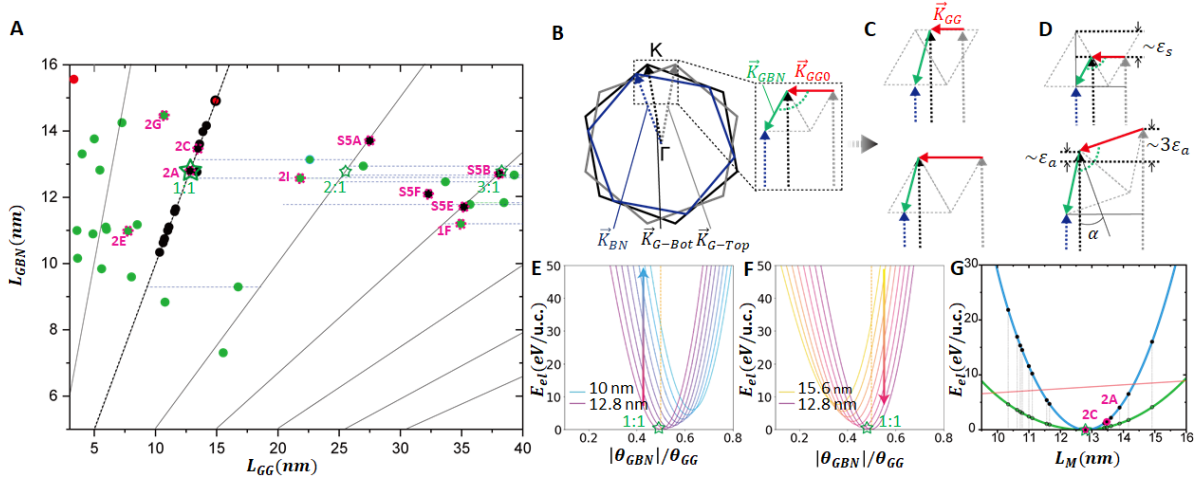

Fig. 3. **Mechanism of moiré self-alignment.** (A) $L_{GBN}$ versus $L_{GG}$ phase diagram classifying the double moiré patterns. Data points corresponding to Fig. 1, Fig. 2 and Fig. S5 are labeled in pink. The black dashed lines represent $L_{GG} = \frac{n}{m} L_{GBN}$; $n, m \in \mathbb{Z}^+$, and the rigid lattice $m:n$ commensurations are marked by a green star. The observed 1:1 commensurate patterns coincides with the $L_{GG} = L_{GBN}$ line, but many are quite far from the rigid lattice commensuration. Similarly, most observed 1:2 and 1:3 commensurate patterns don't coincide with the rigid lattice commensuration points. (B) Schematic drawings of reciprocal lattice vectors of top and bottom graphene layers ($\vec{K}_{G-Top}$, $\vec{K}_{G-Bot}$) and hBN ($\vec{K}_{hBN}$). The zoomed-in inset shows that the 1:1 at $120°$ commensuration occurs when the moiré reciprocal lattice vectors of GG and GBN, $\vec{K}_{GG}$ (red arrow) and $\vec{K}_{GBN}$ (green arrow) respectively, form a rhombus with inner angles $\varphi = 120°$. (C) Schematic drawing of incommensurate cases with $\theta_{GBN} \neq 1.1°$ (top panel) and $|\theta_{GBN}| \neq \frac{\theta_{GG}}{2}$ (bottom panel). (D) Schematic drawings of self-alignment with symmetric (top) or asymmetric (bottom) strain, $\varepsilon_s$ and $\varepsilon_a$, as described in the text and ref. (36). The tilt angle, $\alpha$, is defined by $\tan \alpha = \left( \theta_{GBN} - \frac{1}{2} \theta_{GG} \right) * \frac{|\vec{K}_{GBN}|}{\frac{\sqrt{3}}{2}|\vec{K}_{GG}|}$. (See discussion of 60° case in SI) (E) & (F) Calculated elastic energy of alignment versus twist-angle ratio for 1:1 commensuration at $120°$ for several moiré wavelength $L_M < 12.4 nm$ ((E)) and $L_M > 12.4 nm$ ((F)). (G) Calculated lowest elastic energy for self-aligned commensuration (blue curve for 60°; green curve for 120°) as a function of $L_M$. The experimentally observed $L_M$ values are estimated assuming they are either 60° or 120° commensuration and plotted with solid and empty circles, respectively. The pink curve marks the estimated energy limit for self-alignment (36). We find that all observed points can be energetically favorable for 120° while some moiré lattices with $L_M$ being close to $L_{M0}$ can stabilize for both 60° and 120° commensuration.

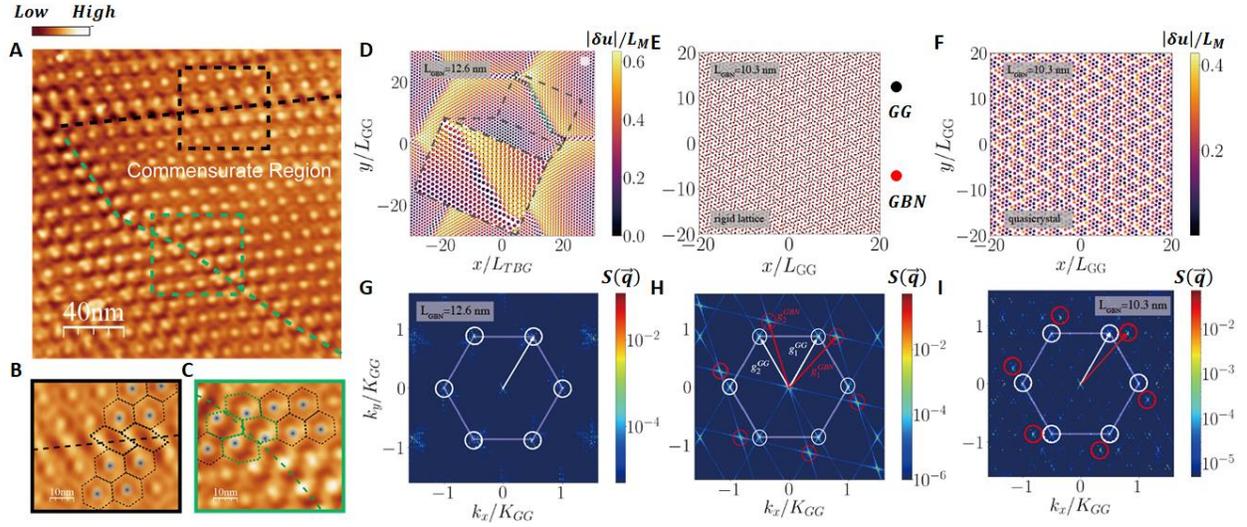

Fig. 4. **Imaging Moiré Self-alignment induced domain and domain boundaries.** (A) STM topography of commensurate double moiré patterns ($L_M = 12.1\ nm$) where 2 types of domain boundaries are observed between different commensurately stacked domains. (B) & (C) Zoom-in of the (B) "tensile" and (C) "shear" domain boundaries from the purple and green rectangle regions in A. The two types of domain boundaries are marked with purple and green dashed lines, respectively. GBN unit cells are outlined by the dashed hexagons and the superlattice distortion is clearly seen near the boundaries. The AA sites are marked by blue spots. The scale bars are 10nm. Tunneling conditions: $V_B$ = -500mV; $V_G$=0; I = 20pA (D-I) Simulated moiré-of-moiré domains and FFT with and without relaxation based on FK-model for $L_{GG} = 12.8\ nm$ ($\theta_{GG}$ =1.10°) and two values of $L_{GBN} = 12.6 nm$ in Fig. (D) and $L_{GBN} = 10.3 nm$ in (E-F). Color scales represent the local magnitude of moiré movement from their initial position before relaxation. (D) Relaxed lattice structure for $L_{GBN} = 12.6\ nm$ forms commensurate domains where the moiré periodicity is locally restored and an emergent moiré-of-moiré structure appears. The inset of (D) shows the two kinds of domain boundaries observed in the experiment: tensile and shear marked by black and green lines respectively. Additionally, the structure factor $S(\vec{q})$ in (G) shows peaks at the moiré periodicity of GG. In this regime, lattice relaxation gives rise to large domains where commensuration is restored with an emergent super-moiré periodicity. Panel (E) shows the interference patter originating from the superposition of the two rigid moiré lattices GG and GBN. After lattice relaxation we find the pattern in (F) characterized by the absence of translational invariance. (H) and (I) Structure factor $S(\vec{q})$ calculated for (E) & (I) respectively. 1st Brillouin zone of GBN is marked by the light grey hexagon. In Fig. (I) additional non commensurate peaks originating from the interaction with BN are shown in red while the GG periodicity in white.

# ACKNOWLEDGMENTS
**Funding:**
Department of Energy grant DOE-FG02-99ER45742 (EYA, XL)
Gordon and Betty Moore Foundation EPiQS initiative grant GBMF9453 (EYA, XL)
Flatiron Institute, a division of the Simons Foundation (DG)
Rutgers University (GL)
Air Force Office of Scientific Research grant No. FA9550-20-1-0136 (JHP)
NSF CAREER grant No. DMR-1941569 (JHP)
Sloan Research Fellowship through the Alfred P. Sloan Foundation (JHP)
NSF CAREER grant DMR-2238895 (JHW)
Aspen Center for Physics where part of this work was performed, which is supported by National Science Foundation grant PHY-1607611 (JHP, JHW)
Elemental Strategy Initiative conducted by MEXT JPMXP0112101001, CREST JPMJCR15F3 , JSPS KAKENHI P20H00354  (KW and TT)

XL would like to express heartfelt thanks to Yuhang Jiang and Jinhai Mao for their patient guidance during the early stages of his training in STM techniques. We thank David Goldhaber-Gordon for insightful discussions and thank Shiang Fang for discussions and collaborations at

## Supplementary Materials

1. **Sample fabrication**

Twisted bilayer graphene aligned to hBN (GG-GBN) devices were made with the tear and stack technique used in our previous work (*1*). A piece of about 30 nm thick hBN flake was exfoliated to SiO$_2$ capped Si substrate and a monolayer graphene flake was exfoliated to polymethyl-methacrylate (PMMA) 950 A11 membrane first. A piece of polydimethylsiloxane (PDMS) stack was added to the back of PMMA membrane to hold the PMMA onto a glass slide and created a bump around the target graphene flake. The glass slide carrying graphene-PMMA-PDMS stack was moved to a homebuilt transfer stage within an Argon filled glovebox where the monolayer graphene was aligned with the hBN crystal by collocating their crystal edges. Limited by the resolution of optical microscope and rotation control stage, we estimate the alignment error of two individual graphene and hBN flakes to at $\pm 1°$. Half of the graphene flake was brought into contact with the hBN surface. The PMMA membrane was then lifted at room temperature leaving the contacted half of graphene on the substrate while the other torn-off half stayed with the PMMA. The substrate supporting the graphene on hBN heterostructure was then rotated by 1°. The torn-off half graphene was then aligned and pressed onto the twisted graphene on hBN. Lifting the PMMA membrane again at room temperature left the second graphene flake which adhered to the first, forming a GG stack with the bottom layer aligned to hBN. Another layer of thin PMMA 950 A6 was then spin coated to the surface and with this PMMA layer over the GG-GBN stack we could optically identify the location of GG on hBN. One electrode window was put onto the PMMA around the GG stack through standard e-beam lithography. The heterostracture was subsequently contacted by an evaporating a Ti/Au (4nm/40nm) electrode onto the PMMA window in high vacuum at room temperature with a homebuilt e-beam evaporation system. The device wae soaked in acetone overnight for liftoff at room temperature. The shape of the electrode was customized to allow for capacitive navigation towards the sample region at low temperature (*2*). Finished devices were annealed in forming gas (10% Hydrogen and 90% Argon) at $300°C$ for more than 24 hours to remove all polymer residue before STM measurements.

2. **STM measurement**

STM and STS measurements were performed on a homebuilt low temperature high vacuum STM with a base temperature T = 4.5 K using an etched tungsten tip (*3*). The tip was prepared on a gold surface at base temperature and calibrated at the nearby monolayer graphene area on the sample with a V-shape dI/dV spectrum for Dirac fermions. The sample was supported on a ~300nm SiO$_2$ layer capping an n-doped Si substrate which serves as a back gate. Gate voltage V$_G$ applied to a doped Si back-gate separated from the sample by the SiO$_2$ and hBN dielectric, while bias voltage V$_B$ (-500mV unless otherwise specified) was applied to the sample to adjust the sample doping and maintaining a constant tunneling current $I = 10 pA$ unless otherwise specified. The STM tip was navigated to the sample area using a STM tip-electrode capacitance sensing technique (*2*). Regions with different moiré wavelengths ($L_{GG}, L_{GBN}$) are found at various locations on two different devices by moving across the sample with coarse piezo motors.

## 3. Identification of stacking order

Fig. S1 shows STM topography images taken at different sample doping and set points of the same region. The triangular GG moiré superlattices characterized by GG AA sites appear higher at full than at empty filling of the flat band by comparing Fig. S1A, 1B with 1C, 1D. This accords with the fact that the wave function of the GG moiré flat band strongly is localized at the AA sites (4). The bottom GBN moiré patterns are no longer visible in Fig. S1B but a broken C3 rotational symmetry could be clearly observed from the distortion of GG AB/BA sites. On the contrary, GBN moiré patterns dominate, and we observe mostly hexagonal patterns in Fig. S1C. From the evolution of contrast, we find GG AA sites overlap with GBN $C_B$ sites thus we can confirm the stacking order of this region is AAB. An AAB unit cell consists of a bright spot surrounded by a hexagon with both GG and GBN moiré patterns having similar contrasts under our choice of parameters: $V_B$ = -500mV; $V_G$ = 0 V. More examples of commensurate GG-GBN double moiré patterns with AAB stacking order at $L_{GG}: L_{GBN} = 1:1$ can be found in Fig. S3.

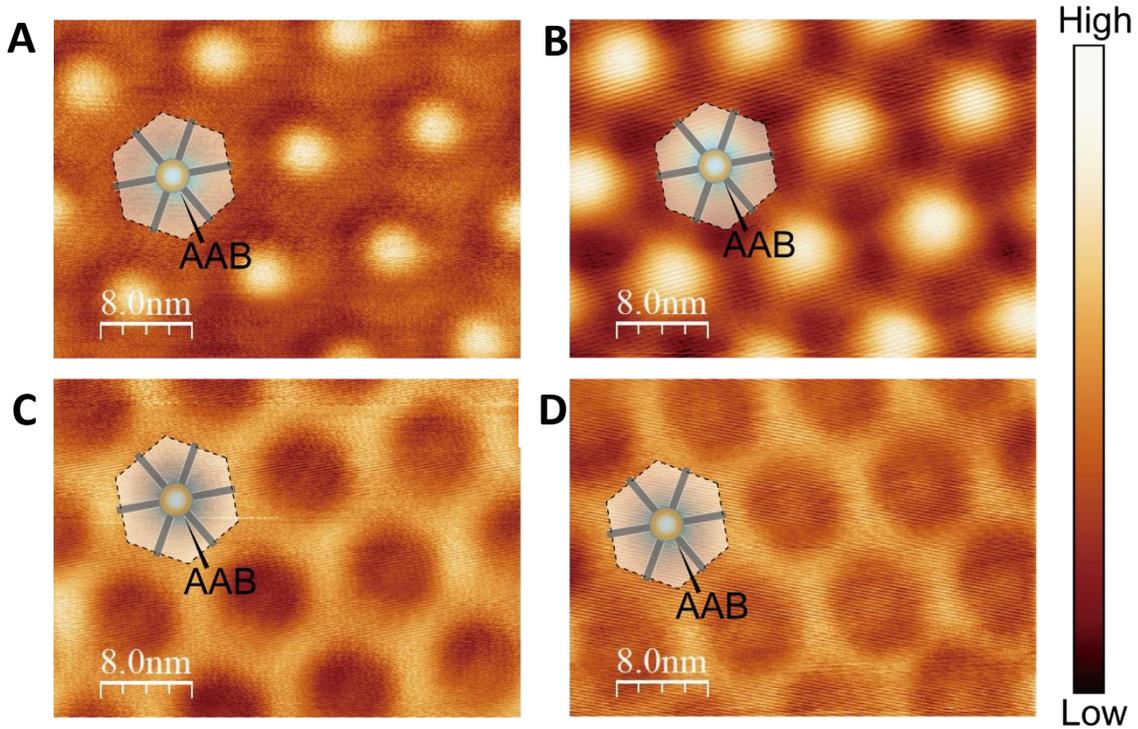

**Fig. S1. Gate and Bias dependence of topographies.** Topography of region with $\theta_{GG} = 1.24°; \theta_{GBN} = 0.79°$ at different $V_B$ and doping: (**A**) $V_B = -500mV$; Full Filling ($V_G = 60V$). GG has more contrast than GBN. (**B**) $V_B = -200mV$; Full Filling ($V_G = 60V$). GG dominates. (**C**) $V_B = -200mV$; Empty Filling ($V_G = -55V$). GBN dominates. (**D**) $V_B = -500mV$; Empty Filling ($V_G = -55V$). GBN has more contrast than GG. A schematic drawing of one AAB unit cell is overlapping with one unit cell in each topography as a guide of eye. Tunneling current: I = 20pA. Topography of the same region taken with the typical measurement parameter used in this work: $V_B = -500mV$; I = 20pA; $V_G = 0V$ is presented in Fig. S3F.

## 4. Stacking order stability

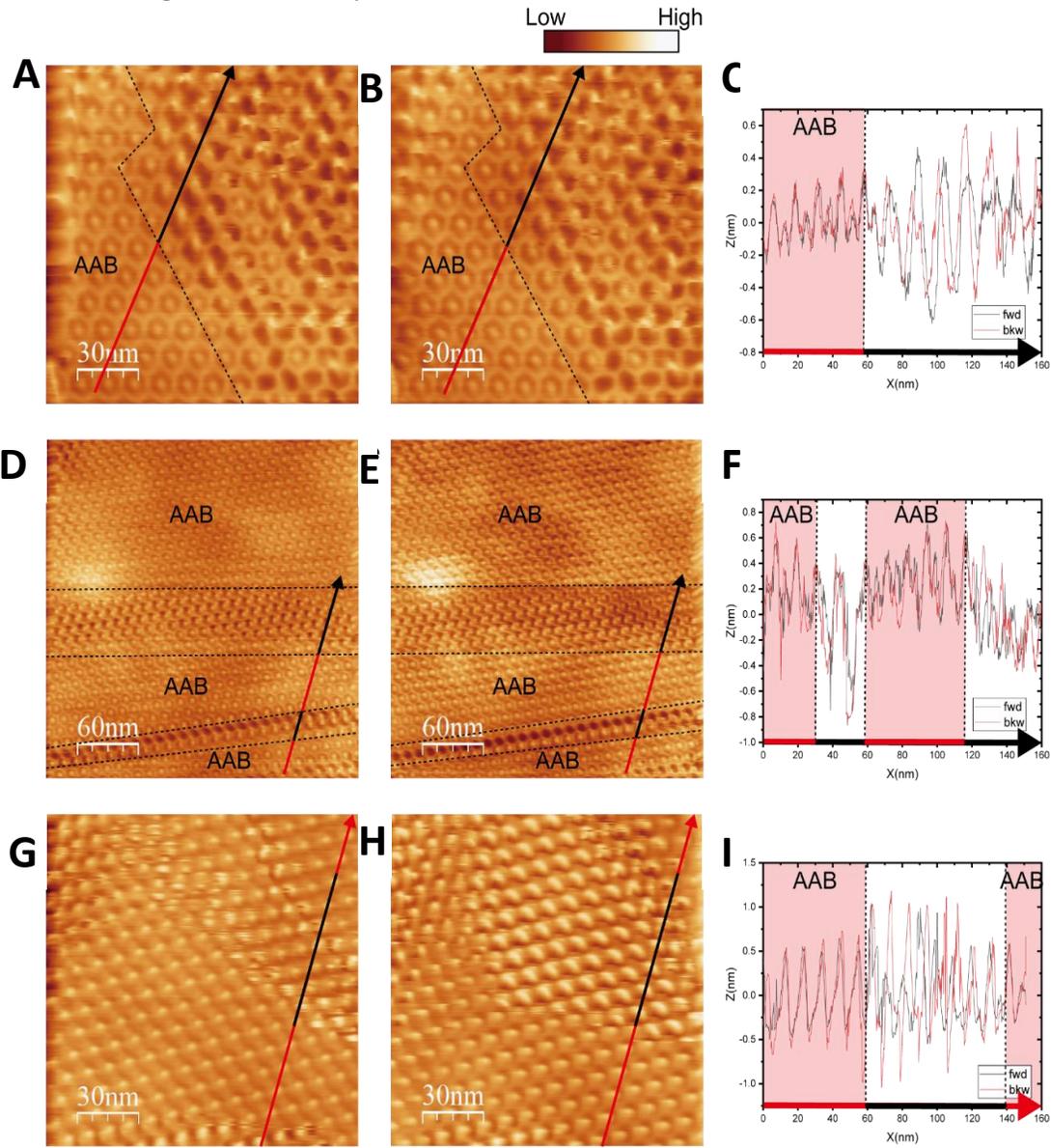

Fig. S2. **Comparing forward and backward STM scans.** (**A**) Forward (left to right) topography scan of region with $L_M = 13.5\ nm$ ($\theta_{TBG} = 1.05°; \theta_{G-hBN} = 0.30°$ for rigid lattices) (**B**) Backward (right to left) scan of region in Fig S3A. (**C**) Linecut of height profile along the arrow in Fig S2A and S2B. (**D**) Forward (left to right) topography scan of region with $L_M = 11.0\ nm$ ($\theta_{TBG} = 1.28°; \theta_{G-hBN} = 0.80°$) that is large enough to include the two kinds of domain boundaries. (**E**) Backward (right to left) scan of the same region in Fig S2D. (**F**) Linecut of height profile along the arrow in Fig S2D and S2E. (**G**) Forward (left to right) topography scan of region with $L_M = 10.3\ nm$ ($\theta_{TBG} = 1.36°; \theta_{G-hBN} = 0.93°$). (**H**) Backward (right to left) scan of the same region in Fig S2G. (**I**) Linecut of height profile along the arrow in Fig S2G and S2H. AAB stacked regions are labeled in each figure.

The forward and backward scans presented in Fig. S2 are the same within the AAB stacked domains but are different at the non-AAB regions. This is especially clear in the comparison of line cuts in Fig. S2C, S2F, S2I where AAB stacked regions match in the forward and backward profiles while non-AAB stacked regions mismatch in terms of both the phases and magnitudes. This proves that non-AAB stacked regions are less stable under external perturbations such as scanning probes or mechanically stacking other layers. AAB is likely preferred among the three of the C3 symmetric stacking order, AAB, AAC and AAN.

## 5. Topography of commensurate stacking

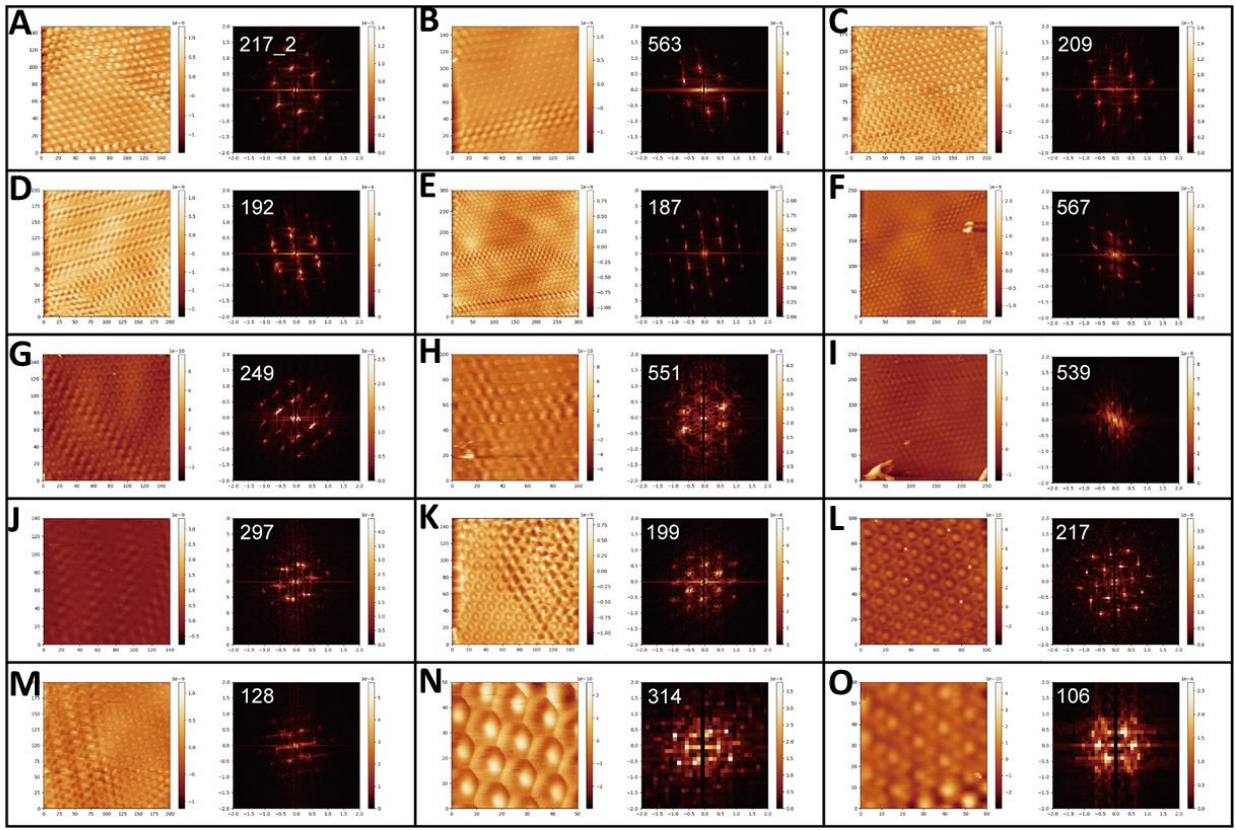

Fig. S3. **STM topographies of $L_{GG} = L_{GBN}$ Commensurate cases with the corresponding FFT.** The file name of each group is marked in the FFT. The peaks at the second Brillion zone have high intensity and are used as indicators for GG-GBN double moiré patterns. Measured moiré wavelengths and measurement errors for each figure are in Table S1 The unit of topography coordinates is $nm$. The unit of FFT coordinates is $\pi/nm$ and range of FFT is $\pm 2\pi/nm$. The topographies are in ascending order of $L_{GG}$. The measured wavelengths are shown in Table S1

The examples in Fig. S3 show AAB stacking are observed to be the only stacking order in all 1:1 commensurate regions up to $\theta_{GG} \approx 1.4°$ in GG-GBN. We find these commensurate AAB regions are either separated by domain boundaries or adjacent to a disordered region with various possible stacking orders.

| Fig. S3 | $L_{GG}$ (nm) | $L_{GBN}$ (nm) | Main text and SM Figure |
|---|---|---|---|
| A | 10.3±0.3 | 10.3±0.3 | S2G |
| B | 10.6±0.3 | 10.6±0.3 | |
| C | 10.7±0.2 | 10.7±0.2 | |
| D | 10.8±0.3 | 8.8±0.2 | |
| E | 11±0.2 | 11±0.2 | S2D |
| F | 11.1±0.2 | 11.1±0.2 | 2A |
| G | 11.6±0.4 | 11.6±0.4 | |
| H | 11.7±0.6 | 11.7±0.6 | |
| I | 12.8±0.3 | 12.8±0.3 | 2C/4A |
| J | 13.4±0.6 | 12.8±0.5 | |
| K | 13.5±0.5 | 13.5±0.5 | S2A |
| L | 13.6±0.8 | 13.6±0.8 | |
| M | 13.9±0.4 | 14±0.4 | |
| N | 14±2 | 14±2 | |
| O | 15±2 | 15±2 | |

Table S1. **Measured moiré wavelengths from FFTs in Fig. S3.** Since we measure the wavelength of commensurate case at the second Brillouin zone, half the size of the specific pixel at the measured wavelengths translate to real space as $error = \frac{L_M}{(2 - \frac{\sqrt{3}}{8s}L_M)} - \frac{L_M}{(2 + \frac{\sqrt{3}}{8s}L_M)}$; $L_M = L_{GG}$ or $L_{GBN}$ and $s$ is the width of the corresponding topography image.

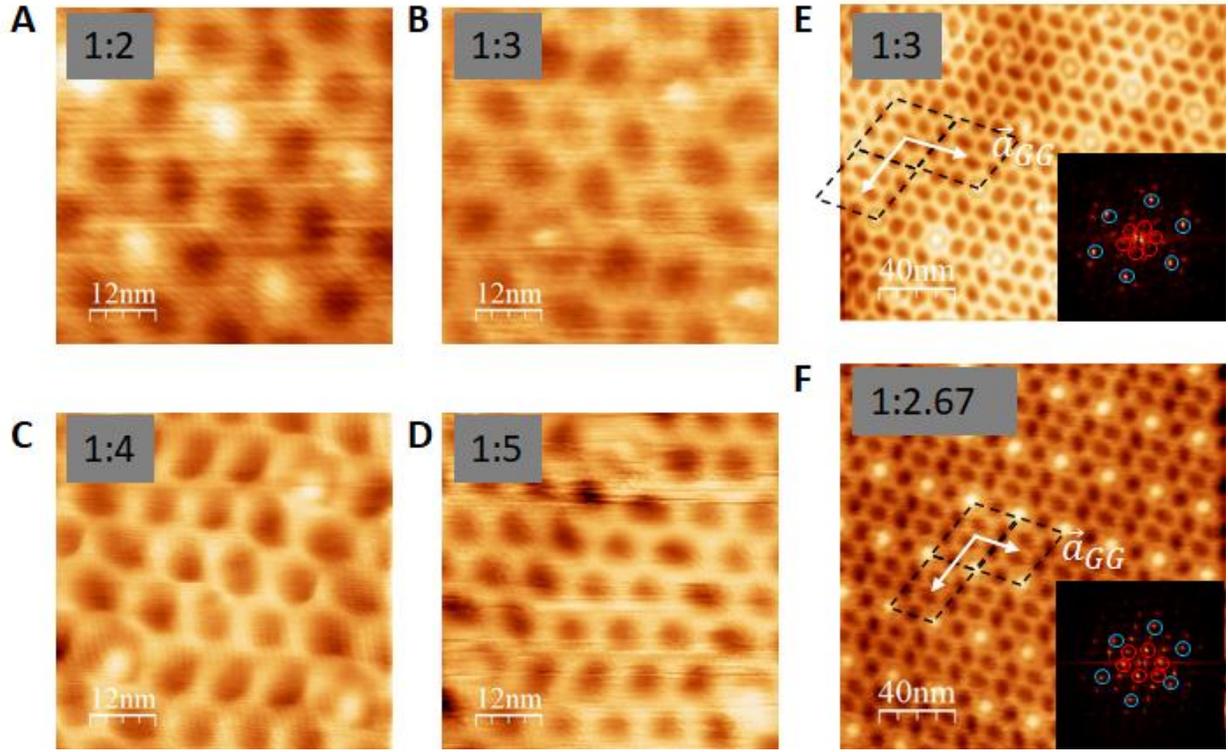

Fig. S4. **Self-aligned commensurate structures other than 1:1 at small $\theta_{GG} < 1°$.** The self-alignment leads to many locally commensurate cases at small twist angles in the strong moiré alignment regime for $L_{GG}:L_{GBN} = 1:2, 1:3, 1:4, 1:5$ in **(A)** $L_{GG} = 27.5\ nm; L_{GBN} = 13.7\ nm$. **(B)** $L_{GG} = 38.1\ nm; L_{GBN} = 12.7\ nm$. **(C)** $L_{GG} = 44.5\ nm; L_{GBN} = 11.1\ nm$. **(D)** $L_{GG} = 50.3\ nm; L_{GBN} = 10.1\ nm$. respectively. **(E)** Large commensurate domains are observed for the 1:3 case with $L_{GG} = 35.2\ nm; L_{GBN} = 11.7\ nm$. The inset is FFT of this topography, the window size of FFT is 0.4 by 0.4 (1/nm). **(F)** Large commensurate domains where the GG moiré pattern is subject to uniaxial strain and clear distorted moiré patterns are observed. Here the average moiré wavelength is $L_{GG} = 32.3.\ nm; L_{GBN} = 12.1\ nm$. The translational symmetry is preserved in this case making it commensurate. The primitive superlattice vectors are marked by white arrows and primitive cells are outlined by dashed diamonds. The insets in **(E)** and **(F)** are FFTs of the corresponding topography, the window size of FFT is 0.4 by 0.4 (1/nm). Blue and red circles in FFTs highlight the peak for GBN and GG wavevectors, respectively.

GG-GBN topography images for $\theta_{GG} < 1°$ self-aligned other than 1:1 commensurate structures are shown in Fig. S4. We note that most GG AA sites align to the nearest GBN $C_B$ sites, forming the preferred local AAB stacking. The spacing between each AAB sites ($L_{GG}$) in these topographies is an integer multiple of $L_{GBN}$. This differs from the smooth wavelength transitions observed in previous works arising from strain and twist angle gradients (*5, 6*). This further confirms that AAB is the preferred stacking and can form spontaneously as discussed in the next section.

## 6. Topography of Quasicrystals

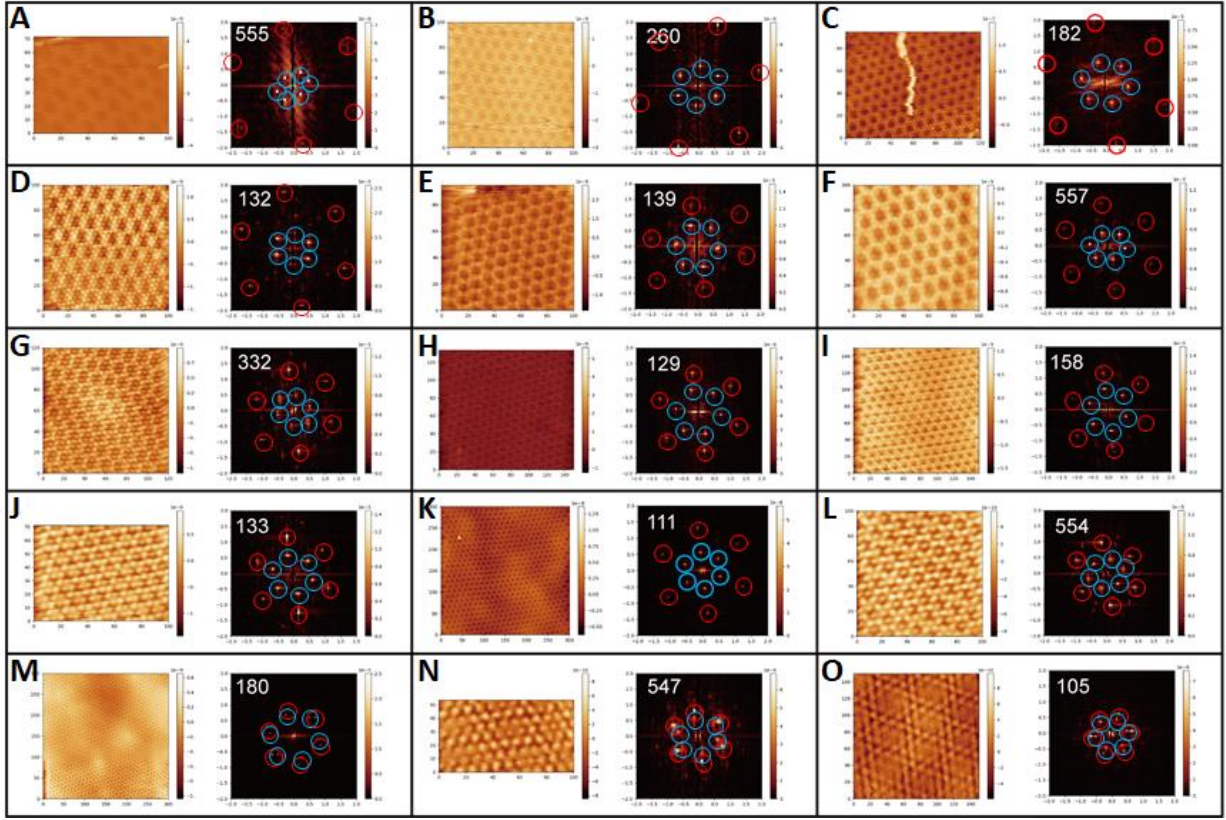

Fig. S5. **STM topographies of Incommensurate cases without and with moiré-of-moiré patterns (IC and IC-MoM respectively) and their corresponding FFT in the weak moiré interaction regime.** The file name of each group is marked in the FFT. Measured moiré wavelengths and measurement errors for each figure are in Table S2. Topographies shown in **(M), (N), (O)** show very close periodicities in FFT. Topography coordinate units are $nm$. FFT coordinate units are $\pi/nm$ and the FFT range is $\pm 2\pi/nm$. The topographies are in ascending order of $L_{GG}$. Blue and red circles in FFTs highlight the Bragg peaks of GBN and GG respectively.

| Fig. S5 | $L_{GG}$ (nm) | $L_{GBN}$ (nm) | $L_C$ (nm) | Main text Figure |
|---|---|---|---|---|
| A | 3.33±0.05 | 16±1 | 656 | |
| B | 3.59±0.06 | 11.0±0.5 | 3091 | |
| C | 3.64±0.05 | 10.2±0.4 | 928 | |
| D | 4.01±0.07 | 13.3±0.8 | 798 | |
| E | 4.9±0.1 | 10.9±0.5 | 534 | |
| F | 5±0.1 | 13.8±0.8 | 345 | |
| G | 5.5±0.1 | 12.8±0.6 | 704 | |
| H | 5.61±0.09 | 9.8±0.3 | 5498 | |
| I | 6.0±0.1 | 11.1±0.4 | 222 | |
| J | 6.0±0.2 | 11±0.5 | 66 | |
| K | 7.1±0.1 | 10.9±0.2 | 774 | 2E |
| L | 7.3±0.2 | 14.2±0.9 | 1037 | |
| M | 8.04±0.09 | 9.6±0.1 | 643 | |
| N | 8.5±0.3 | 11.2±0.5 | 952 | |
| O | 10.7±0.3 | 14.5±0.6 | 1552 | 2G |

**Table S2. Measured moiré wavelengths from FFTs in Fig. S5.** The size of the pixel at the measured wavelength translate to real space as $error = \frac{L_M}{(1-\frac{\sqrt{3}}{4s}L_M)} - \frac{L_M}{(1+\frac{\sqrt{3}}{4s}L_M)}; L_M = L_{GG}$ or $L_{GBN}$ and $s$ is the width of the corresponding topography image. We find the closest rational number $\frac{n}{m}$ with $n$ and $m$ being integers to the ratio $\frac{L_{GG}}{L_{GBN}}$ for each cases in the table. We then estimate the possible commensurate length based on the measured wavelength using $L_C \approx L_{GG} * n \approx L_{GBN} * m$. Patterns with $L_C$ larger than the sample size are classified as quasicrystals. We note that there is a separation of length scales with $L_C$ being at least one order of magnitude larger than $L_{GG}$ or $L_{GBN}$ for all cases studied here.

## 7. Topographies of Incommensurate lattices with strong moiré interaction

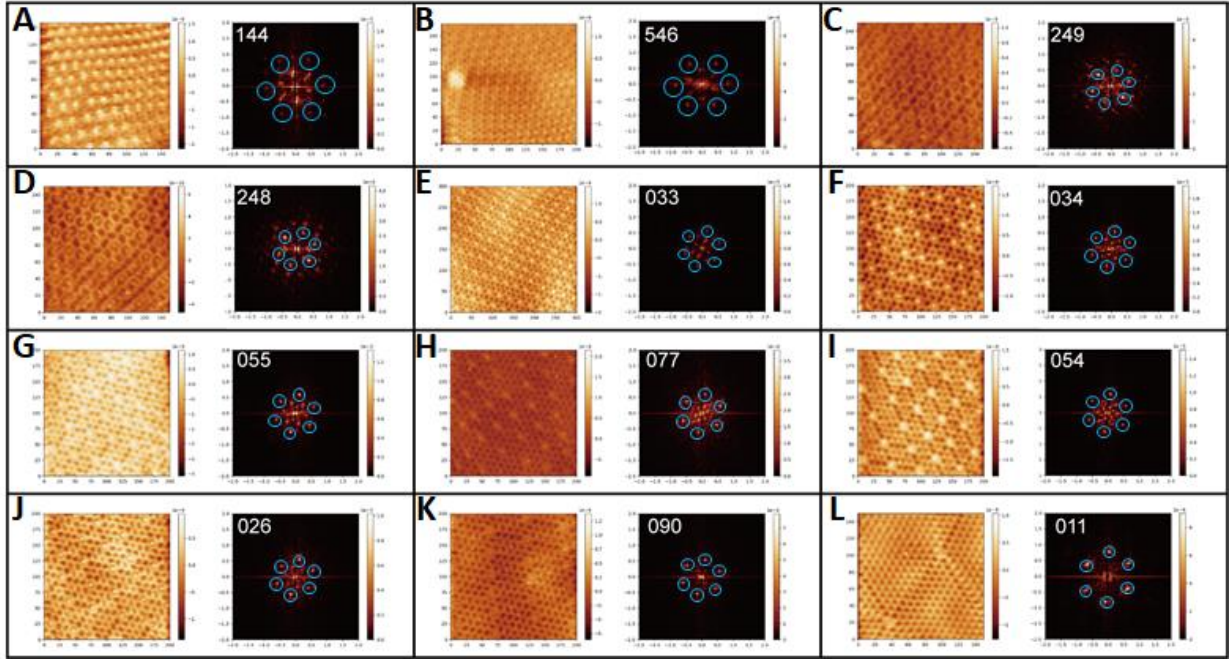

Fig. S6. **STM topographies of Incommensurate cases with strong moiré interaction and their the corresponding FFT.** The file name of each group is marked in the FFT. Measured moiré wavelengths and measurement errors for each figure are in Table S3. The unit of topography coordinates is $nm$. The unit of FFT coordinates is $\pi/nm$ and range of FFT is $\pm 2\pi/nm$. The topographies are in ascending order of $L_{GG}$. Blue circles highlight the Bragg peaks corresponding to GBN. Due to the strong self-alignment in these data, the FFT peaks for GG wavevectors are not well defined and we typically observe a group of 6 clouds near the center of each frame for GG.

| Fig. S6 | $L_{GG}$ (nm) | $L_{GBN}$ (nm) | $L_C$ (nm) | Main text Figure |
|---|---|---|---|---|
| A | 15.5±0.7 | 7.3±0.2 | 1132 | |
| B | 16.5±0.6 | 9.3±0.2 | 512 | |
| C | 22±1 | 12.6±0.5 | 1386 | 2I |
| D | 23±1 | 13.1±0.5 | 3013 | |
| E | 27±1 | 12.9±0.2 | 1161 | |
| F | 34±2 | 12.5±0.3 | 850 | |
| G | 35±3 | 11.2±0.3 | 280 | 1F |
| H | 36±3 | 11.8±0.3 | 2124 | |
| I | 38±3 | 11.8±0.3 | 2242 | |
| J | 39±3 | 12.7±0.3 | 4953 | |
| K | 61±8 | 12.6±0.3 | 3843 | |
| L | NA | 9.6±0.3 | NA | |

Table S3. Measured moiré wavelengths from FFTs in Fig. S6. The size of the specific pixel at the measured wavelengths translate to real space as $error = \frac{L_M}{(1-\frac{\sqrt{3}}{4s}L_M)} - \frac{L_M}{(1+\frac{\sqrt{3}}{4s}L_M)}; L_M = L_{GG}$ or $L_{GBN}$ and $s$ is the width of the corresponding topography image. We find the closest rational number $\frac{n}{m}$ with $n$ and $m$ being integers to the ratio $\frac{L_{GG}}{L_{GBN}}$ for each cases in the table. We then estimate the possible commensurate length based on the measured wavelength using $L_C \approx LL_{GG} * n \approx L_{GBN} * m$. Patterns with $L_c$ larger than the sample size are classified as quasicrystals. We note that there is a separation of length scales with $L_C$ being at least one order of magnitude larger than $L_{GG}$ or $L_{GBN}$ for all cases studied here.

## 8. Derivation of the moiré self-alignment model

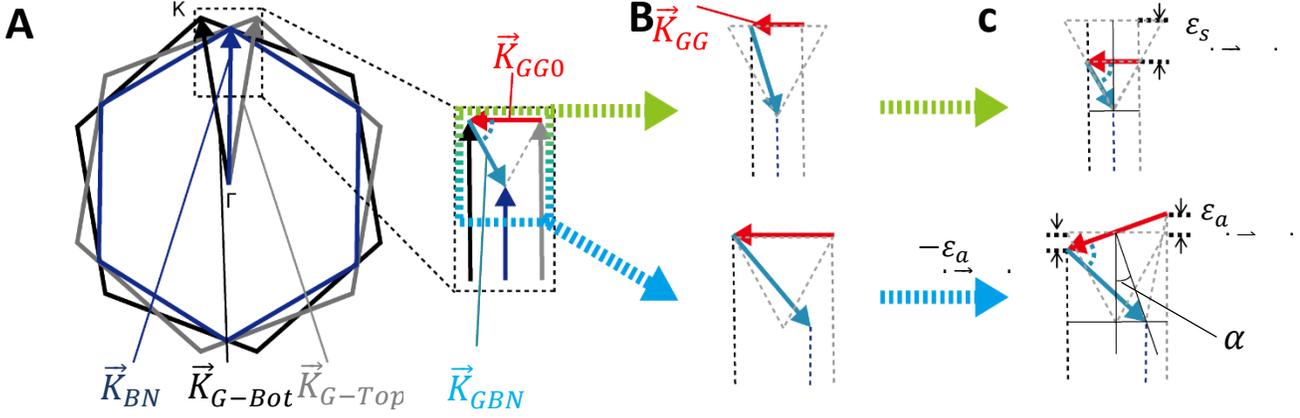

Fig. S7. **Geometric analysis of lattice rescaling for 60° 1:1 commensurate cases. (A)** Schematic drawings of reciprocal lattice vectors of top and bottom graphene layers ($\vec{K}_{G-Top}$, $\vec{K}_{G-Bot}$) and hBN ($\vec{K}_{BN}$). Zoom-in shows the 1:1 rigid lattice commensurate condition for 60° commensuration. **(B)** Schematic drawing of incommensurate cases with $\theta_{GBN} \neq 1.1°$ (top panel) and $|\theta_{GBN}| \neq \frac{\theta_{GG}}{2}$ (bottom panel). **(C)** Schematic drawings of self-alignment with symmetrical ($\varepsilon_s$, top panel) or asymmetrical ($\varepsilon_a$, bottom panel) lattice rescaling. Here $\alpha$ is the tilt angle defined by the asymmetrical lattice rescaling as discussed in the main text.

Since the GG-GBN samples are prepared with monolayer graphene transferred to a flake of bulk hBN with thickness > 30 nm, we assume that only the graphene layers stretch/shrink during the moiré self-alignment process. We also did not observe any signature of strain in the top hBN layer which would induce a third set of moiré patterns.

We first solve for the required symmetric lattice strain $\varepsilon_s$ for any combination of $\theta_{GG}$ and $\theta_{GBN}$ as long as $|\theta_{GBN}| = \theta_{GG}/2$ is satisfied. $\varepsilon_s = \frac{K_G}{K_{G'}} - 1 = \frac{a_{G'}}{a_G} - 1$ is the symmetrical lattice strain. $K_G$ is the reciprocal lattice constant of graphene; $K_{G'}$ is the graphene reciprocal lattice constant after symmetric relaxation; $K_{BN}$ is the reciprocal lattice constant of hBN; $\theta_{GG}$ is the GG twist angle; $\theta_{GBN}$ is the GBN twist angle; $\delta = \frac{K_G}{K_{BN}} - 1 = \frac{a_{BN}}{a_G} - 1$ is the lattice mismatch of G and hBN. Since $|\vec{K}_{GG}| = |\vec{K}_{GBN}|$; $\angle(\vec{K}_{GG}, \vec{K}_{GBN}) = 60°$ or $120°$ when GG and GBN moiré patterns are 1:1 commensurate, we can solve the equations: $\frac{2}{\sqrt{3}}(K_{G'}\cos\left(\frac{\theta_{GG}}{2}\right) - K_{BN}) = 2K_{G'} * \sin\left(\frac{\theta_{GG}}{2}\right)$; $K_{G'} = \frac{K_G}{1+\varepsilon_s} = \frac{(1+\delta)K_{BN}}{1+\varepsilon_s}$. This gives: $\varepsilon_s = \left(\cos\left(\frac{\theta_{GG}}{2}\right) - \sqrt{3}\sin\left(\frac{\theta_{GG}}{2}\right)\right)(1+\delta) - 1$. The 1:1 alignment happens near $\theta_{GG}$ and $\theta_{GBN}$ around 1° so we simplify this formula with small angle approximation:

$$\varepsilon_s = \delta - \frac{\sqrt{3}}{2}\theta_{GG}. \tag{1}$$

This formula is identical for both 60° and 120° 1:1 commensuration.

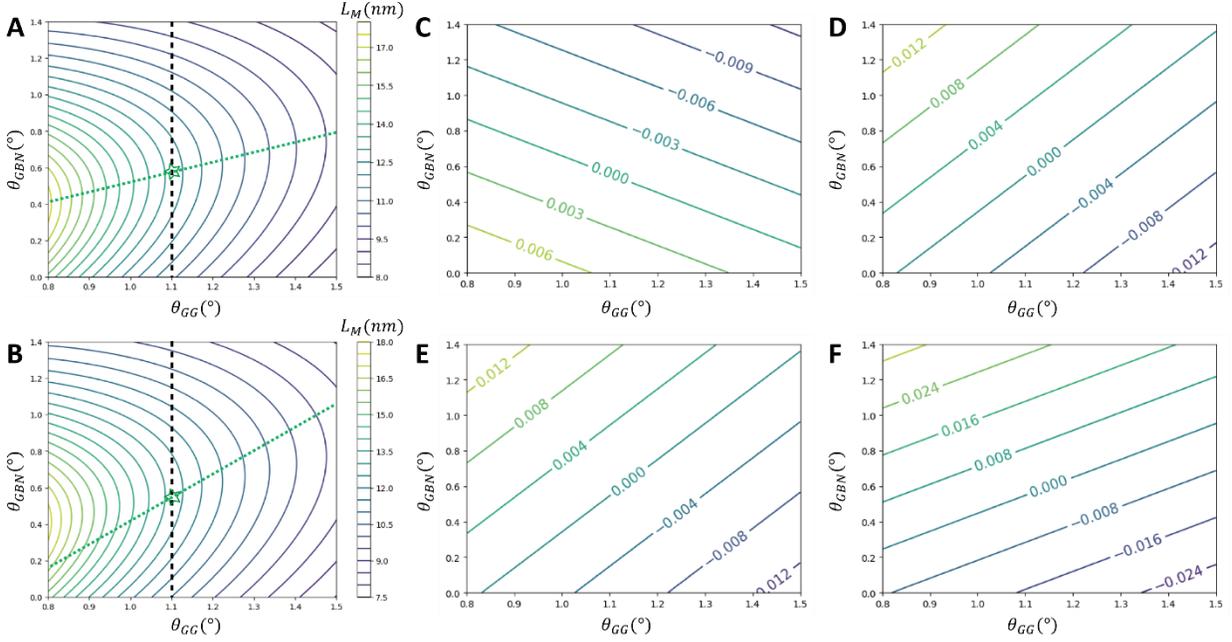

Fig. S8. **Simulated twist angle dependence of moiré wavelength and strain for 1:1 commensurate cases (60° and 120°).** Contour plot of moiré wavelength $L_M$ after self-alignment vs. $\theta_{GG}(°)$ and $\theta_{GBN}(°)$ for the **(A)** 60° commensuration and **(B)** 120° commensuration. The green line marks the lowest energy cost twist angle combination for a certain $L_M$. The green stars marks the rigid lattice commensurate conditions. **(C)** Contour plot of bottom layer strain $\varepsilon_{bot}$ versus twist angles for the 60° commensuration. **(D)** Contour plot of top layer strain $\varepsilon_{top}$ versus twist angles for the 60° commensuration. **(E)** Contour plot of bottom layer strain wavelength $\varepsilon_{bot}$ versus twist angles for the 120° commensuration case. **(F)** Contour plot of top layer strain $\varepsilon_{top}$ for $\theta_{GG}(°)$ and $\theta_{GBN}(°)$ for the 120° commensuration.

Asymmetric lattice strain $\varepsilon_a$ can be determined from the geometrical relation in the lower panel of Fig 3D or Fig. S7C when applying small angle approximations. We have:

$$\varepsilon_a = \frac{\theta_{GG}}{2\sqrt{3}} - \frac{\theta_{GBN}}{\sqrt{3}}. \tag{2}$$

We also show by comparing Fig 3C, 3D with Fig. S7B, S7C that 120° and 60° commensuration are geometrically similar where the top and bottom lattice distortions are the combination of symmetric and asymmetric lattice strain terms $\varepsilon_s$ and $\varepsilon_a$. For 120° cases, the asymmetric lattice strain on top layers is the asymmetric lattice strain of bottom layer magnified by three times: $\varepsilon_{bot} = \varepsilon_s + \varepsilon_a$; $\varepsilon_{top} = \varepsilon_s + 3\varepsilon_a$. While for 60° cases, the asymmetric lattice strain is inverted for top and bottom graphene layers such that $\varepsilon_{bot} = \varepsilon_s + \varepsilon_a$; $\varepsilon_{top} = \varepsilon_s - \varepsilon_a$. Since 120° and 60° commensuration cases are indistinguishable under STM topography and they all host AAB,

AAN and AAC stacking orders, the topographies studied may include both the cases. We have not observed any region with stable 90° commensuration which would produce two separate periodicities ($L_{GG}: L_{GBN} = \sqrt{3}: 2$) in Fourier transformations.

We predict the commensurate moiré wavelength for all twist angle combinations with the relaxed atomic lattices:

$$L_M = \frac{(1 + 2|\varepsilon_a|)(1 + \varepsilon_s - |\varepsilon_a|)a}{\sqrt{2(1 + 2|\varepsilon_a|)(1 - \cos(\theta_{GG})) + 4|\varepsilon_a|^2}}. \tag{3}$$

$a = 0.246 nm$ is the lattice constant of graphene. We plotted the twist angle ($\theta_{GG}$ and $\theta_{GBN}$) dependence of $L_M$ in Fig. S8A for 60° 1:1 commensurate case and in S8B for 120° cases. With $\varepsilon_s$ and $\varepsilon_a$ derived as a function of $\theta_{GG}$ and $\theta_{GBN}$, we simulated the top layer strain ($\varepsilon_{bot}$) and bottom layer strain ($\varepsilon_{top}$) dependence of twist angles in Fig. S8C and S8D for 60° 1:1 commensurate cases and in Fig. S8E and S8F for 120° 1:1 commensurate cases.

### 9. The moiré-of-moiré patterns from GG and GBN moiré patterns

The moiré-of- moiré patterns formed between GG and GBN moiré patterns is an indicator of their incommensuratbility. The moiré-of- moiré wavelength $L_{MM}$ goes to infinite when GG and GBN patterns are fully commensurate. The GG reciprocal moiré wavevectors $\vec{K}_{GG}$ in perpendicular to graphene reciprocal vectors $\vec{K}_G$ under small angle approximation. The angle between $\vec{K}_{GG}$ and $\vec{K}_{GBN}$ is thus $\Delta\theta_{MM} \approx atan\frac{\delta}{(\delta+1)\theta_{GBN}}$ where $\delta$ is the atomic lattice mismatch between graphene and hBN. With $L_{MM} = \frac{(1+\delta_M)L_{GG}}{\sqrt{2(1+\delta_M)(1-\cos(\Delta\theta_{MM}))+\delta_M^2}}$, we calculate the moiré-of-moiré wavelength for all $L_{GG}$ and $L_{GBN}$ considering the lattices are rigid. This gives the plot in Fig S9A. The $L_{MM}$ characterize the incommensurability near the rigid 1:1 commensurate condition. We then estimate the separation of scales between $L_{MM}$ and moiré length scales ($L_{GG}$ and $L_{GBN}$) in Fig S9B.

### 10. Derivation and simulation of the elastic energy in self-aligned moiré patterns

Graphene lattices as well as the GG moiré pattern formed between them preserved the $C_6$ rotational symmetry. We can deduce the elastic potential energy density for strained six-fold symmetric 2D system:

$$\Delta E_{el} = \frac{K}{2}(\varepsilon_{xx} + \varepsilon_{yy})^2 + \mu\left(\left(\frac{(\varepsilon_{xx} - \varepsilon_{yy})^2}{2} + 2\varepsilon_{xy}^2\right)\right). \tag{4}$$

$K$ is the 2D bulk modulus and $\mu$ is the shear modulus. $u_{ij}$ is the strain tensor.

For homogeneous stretching/compressing of six-fold symmetric system we have $\varepsilon_{xx} = \varepsilon_{yy} = \varepsilon_{\text{top}}$ or $\varepsilon_{\text{bot}}$; $\varepsilon_{xy} = 0$, thus the elastic energy density for one strained graphene layer is: $\Delta E_{el} = 2K\varepsilon_{\text{top/bot}}^2$. Assuming the strain in the top and bottom graphene layers are independent, the elastic energy per GG moiré unit cell within both top and bottom layers is:

$$E_{el-GG} = \Delta E_{el} * \frac{\sqrt{3}}{2} L_M{}^2 = \sqrt{3} t Y L_M{}^2 (\varepsilon_{top}{}^2 + \varepsilon_{bot}{}^2) \tag{5}$$

$Y$ is the Young's modulus and $t$ is the thickness of the 2D material. Considering $Y \sim 0.9\ TPa$ (7) and $t \sim 0.34$ being the thickness one graphene layer, we calculate $E_{el-GG}$ as a function of $L_M$ using the formulas given above.

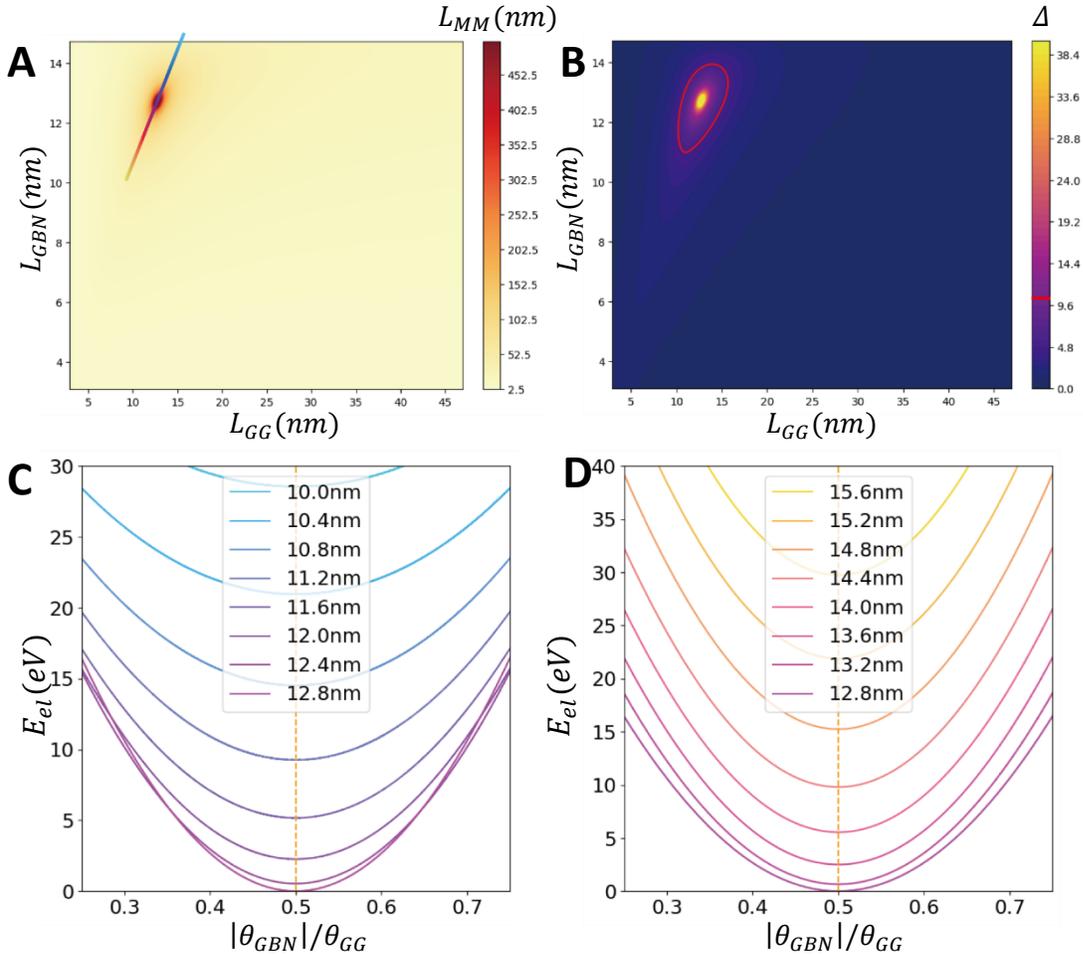

Fig. S9. **Simulated moiré wavelength dependence of moiré-of-moiré patterns and elastic energy for 60° 1:1 commensurate cases.** **(A)** Simulated moiré-of-moiré wavelength $L_{MM}$ as a function of $[L_{GG}, L_{GBN}]$ regarding graphene and hBN lattices are rigid. The $L_{MM}$ is maximized at the theoretical 1:1 commensurate twist angle combination. **(B)** Estimated separation of scale $\Delta = \frac{L_{MM}}{Max[L_{GG},\ L_{GBN}]}$ versus $L_{GG}$ and $L_{GBN}$. We used small angle approximation in calculating $L_{GG}$ and $L_{GBN}$ thus the same plot also applies to 120° commensuration. The red circle is $\Delta = 5$ which defines the commensurate regime in FK-model simulation. Elastic energy ($E_{el}$) at constant wavelength ($L_M$) for different twist angle combinations ($|\theta_{GBN}|/\theta_{GG}$) for 60° 1:1 commensurate cases is plotted in **(C)** for $L_M < 12.8\ nm$. **(D)** for $L_M > 12.8\ nm$.

We simulate the dependence of elastic energy per moiré unit cell $E_{el-GG}$ versus $L_M$ and twist angles. This gives the curve in Fig. 3E and Fig. 3F for 120° commensurate cases. Elastic energy simulations of 60° cases can be found in Fig. S9C and S9D. For 60°, the minimal energy cases are simply at $\varepsilon_a = 0$. Under the small angle approximation, we have:

$$E_{el-GG} \approx 2\sqrt{3}tY \left[\frac{\left(1+\delta - \frac{\sqrt{3}}{2}\theta_{GG}\right)a}{\theta_{GG}}\right]^2 \left(\delta - \frac{\sqrt{3}}{2}\theta_{GG}\right)^2 \approx 2\sqrt{3}tYL_M^2 \left(\delta - \frac{a(1+\delta)}{L_M + \frac{\sqrt{3}}{2}a}\right)^2$$

$$= Yta_G^2 \frac{\sqrt{3}}{2}\left[3\left(\frac{L_M}{L_{M0}} - \cos\alpha\right)^2 + (\sin\alpha)^2\right]. \quad (7)$$

Which is proportional to $L_M^2$ as expected from the discussion in the main text with small angle approximation and $L_M = \frac{\left(1+\delta - \frac{\sqrt{3}}{2}\theta_{GG}\right)a}{\theta_{GG}}$. We plot the blue curve in Fig. 3G using this equation.

Similarly for the 120° case, we derive:

$$E_{el-GG} = Yta_G^2 \frac{\sqrt{3}}{2}\left[3\left(\frac{L_M}{L_{M0}} - \cos\alpha\right)^2 + 3(\cos\alpha)^2 - 2\sqrt{3}\left(\frac{L_M}{L_{M0}} - \cos\alpha\right)\cos\alpha + 1\right]. \quad (8)$$

which leads to the green curve in Fig. 3G.

The preferred stacking of GG-GBN is AA aligned to $C_B$ (AAB). This lowers the local van der Waals (vdW) energy per atom by up to $0.018\ eV$ compared to non-aligned AA sites. The estimated vdW energy gain of AAB aligned GG AA sites is $U \approx p * 0.010 eV$ (8), where $p$ is the number of atoms per AA site. The experimentally measured diameter of GG AA sites ($d_{AA}$) (9) is from 4.6 to 5.4 nm for $\theta_{GG}$ from 1.4° to 0.94° or experimentally observed $L_M$ ranging from 10 to 15 nm. We calculate the number of atoms in an AA site using:

$$p = \frac{\pi d_{AA}^2}{\sqrt{3} * a^2}. \quad (9)$$

This gives the range of $U$ to be from $6.4 eV$ to $8.8 eV$ for $L_M = 10\ nm$ to 15 nm. Connecting these two points gives the pink line in Fig 3G as an upper bound estimate of the self-alignment energy.

The elastic energy cost per aligned moiré region is balanced by the Van der Waals energy gain of AAB stacking, U, so that self-alignment will occur for $U > E_{el}$. We obtain a rough estimate of U based on the area of the moiré unit cells. An estimate of the experimental upper bound of stability can be expressed in terms of $E_B$, the vdW energy gain per unit cell in the AAB configuration compared to surrounding non-aligned sites. We obtain $E_B$ ~53 meV per atomic unit cell for the experimentally observed commensurate cases at 60° and $E_B$ ~10 meV for the 120° configuration.

## 11. Participation Ratio and Strain effects

To analyze the effect of strain on the moiré lattices we calculate and plot the participation ratio $p$ vs. $L_{GBN}/L_{GG}$ where $|\vec{K_{GG_n}}|$ and $|\vec{K_{GBN_n}}|$ are the magnitudes of wavevectors in the three crystallographic directions for GG and GBN respectively.

We calculate the participation ratio using.

$$p = \frac{\sum_{n=1}^{3} \vec{K_{GG_n}}^2 + \sum_{n=1}^{3} \vec{K_{GBN_n}}^2}{\left(\sum_{n=1}^{3} |\vec{K_{GG_n}}| + \sum_{n=1}^{3} |\vec{K_{GBN_n}}|\right)^2}. \tag{15}$$

$p$ is plotted vs. ratio of averaged moiré wavelength $L_{GBN}/L_{GG}$ in Fig S10. The participation ratio characterizes the uniformity of lattice constants in different crystallographic directions. Mismatched $K_{GG}$ and $K_{GBN}$ or strain disorder effectively increases $p$. The participation ratio minimizes near $K_{GG} = K_{GBN}$.

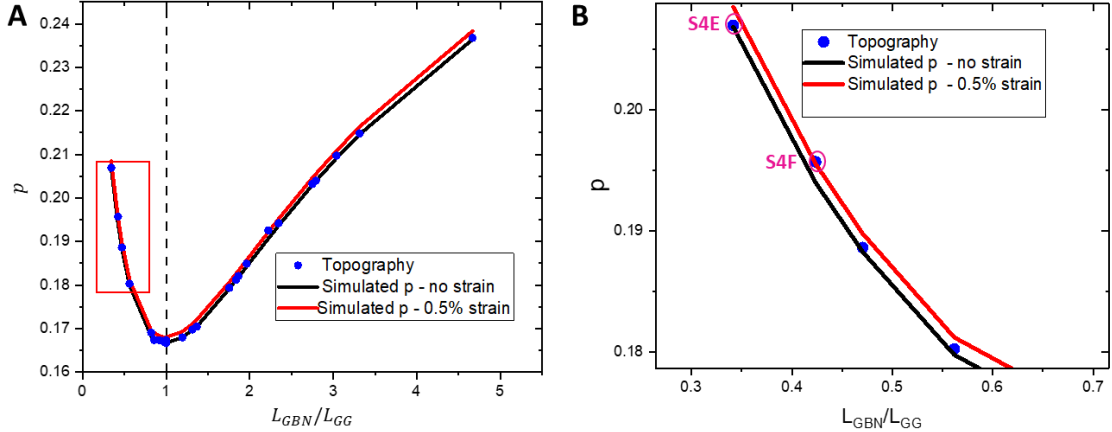

Fig. S10. **Plot of participation ratio for topographies.** **(A)** Plot of participation ($p$) ratio for parts of the experimental data (blue dots). Simulated $p$ for homogeneous lattices versus $L_{GBN}/L_{GG} = K_{GG}/K_{GBN}$ is plotted as the black curve and $p$ with 0.5% uniaxial strain is plotted as the red curve. **(B)** Zoom-in to the red square in **(A)**. Points correspond to large area topographies in Fig S5 are marked by pink circles and labels.

To simulate the strain effect, we first calculate $\vec{k_{G1_n}}$ using graphene lattice constant $a_0 = 0.246nm$ such that $\vec{K_{G1_1}} = \begin{pmatrix} \frac{4\pi}{\sqrt{3}a_0} \\ 0 \end{pmatrix}$

Using formula:

$$\vec{K_{G1_1}}' = \begin{pmatrix} \cos(\theta_s') & \sin(\theta_s') \\ -\sin(\theta_s') & \cos(\theta_s') \end{pmatrix} \begin{pmatrix} \frac{1}{1+\epsilon} & 0 \\ 0 & \frac{1}{1-d\epsilon} \end{pmatrix} \begin{pmatrix} \cos(\theta_s') & -\sin(\theta_s') \\ \sin(\theta_s') & \cos(\theta_s') \end{pmatrix} \vec{K_{G1_1}}. \tag{16}$$

$d = 0.16$ is the poison ratio of graphene. We make $\epsilon = 0.005$; $\theta'_s = 0°$ and recalculate $\overrightarrow{K_{G1_1}}'$; $\overrightarrow{K_{G1_2}}'$ and $\overrightarrow{K_{G1_3}}'$ remain the same to $\overrightarrow{K_{G1_2}}$ and $\overrightarrow{K_{G1_3}}$. We then calculate the moiré lattice vector:

$$\overrightarrow{K_{GG_1}} = \overrightarrow{K_{G2_1}} - \boldsymbol{R}(\theta_{GG}) \cdot \overrightarrow{K_{G1_1}}'; \quad \overrightarrow{K_{GBN_1}} = \overrightarrow{K_{G1_1}}' - \boldsymbol{R}(\theta_{GBN}) \cdot \overrightarrow{K_{hBN_1}}'. \tag{17}$$

$\theta_{TBG}$ and $\theta_{GhBN}$ are derived from: $L = \frac{(1+\delta)a}{\sqrt{2(1+\delta)(1-\cos(\theta))+\delta^2}}$; $L = \frac{12\pi}{\sqrt{3}\sum_{n=1}^{3}(|\overrightarrow{K_n}|)}$. Moiré wavevector $K_n$ used here is the averaged $L_{TBG}$ and $L_{GhBN}$ in three of the crystallographic directions measured from the FFT of topographies in Table S2, S3 and Fig. 2A, C, E, G, I. We apply the 0.5% strain correction to $\overrightarrow{K_{GG_1}}$ and $\overrightarrow{K_{GBN_1}}$ which increased the estimated $p$ to the red curve in Fig S10. We find the strain is low in our sample and most of our data points are within this range, possibly due to the high temperature annealing process of our devices.

### 12. FK-model Simulation

The lattice relaxation dynamics of small angle GG on hBN has been modeled using a 2D Frenkel-Kontorova model (FK-model), which consists of classical particles (that represent the AA sites of GG) that are connected by springs in the presence of the quasiperiodic potential generated by GBN. We argue that the crossover between the commensurate and the incommensurate regime is governed by the competition between the elastic energy of the monolayer and the interlayer GBN vdW interaction.

Considering the two coexisting moiré patterns of GG, and GBN, one immediate question is under what conditions will there be commensurate stacking of the two, resulting in broken $C_{2z}$ symmetry. In general commensurability will occur for all twist angle pairs satisfying $\theta_{GG} \approx \frac{n}{p+q} \times 1.1°$; $\theta_{GBN} \approx \frac{p-q}{p+q} \times 0.55°$ where (n, p, q) is a triplet of coprime integers that characterizes distinct commensurate structures (*10*). Here we focus on twist angle pairs close to the magic angle, $\theta_{GG} \sim 1.05°$, where strong correlation effects are expected (*10, 11*). For rigid graphene and hBN lattices this condition is satisfied for only three angles: $\theta_{GBN} = +0.55°$ (n,p,q) = (1,1,0) and $\theta_{GBN} = -0.55°$ (n,p,q) = (1,0,1), where the moiré lattice vectors of GG and GBN exactly match (1:1), as well as for $\theta_{GBN} \approx 0°$: (n,p,q) = (2,1,1), where the sum of two GG moiré lattice vectors is twice that of the GBN lattice vector (*10*). These double moiré stacking orders break $C_{2z}$ symmetry and can give rise to non-trivial band topology (*10-13*).

We propose a classical 2D FK-model(*14*) to study lattice relaxation of GG on hBN. The model consists of a triangular lattice describing the AA of GG located at the positions:

$$\vec{R}_{n_1,n_2} = n_1 \vec{t}_1 + n_2 \vec{t}_2, \tag{10}$$

where $\vec{t}_{1/2} = L_{GG}(\pm\sqrt{3}/2, 1/2)$. Each lattice site is connected by a spring of elastic constant $K$ to the 6 nearest neighbors giving rise to the elastic energy:

$$E_{el}[\{\vec{u}\}] = \frac{K}{2} \sum_{<i,j>} (\vec{u}_i - \vec{u}_j)^2. \tag{11}$$

where $u_i$ describes the deviation from the lattice position $R_i$ of the atom $i$. The substrate induces a potential on the lattice that breaks one of the mirror symmetries and preserves C3z. Considering the lowest harmonics we have:

$$E_{pot}[\vec{r}] = -U_0 \sum_{i=1}^{3} \cos(\vec{Q}_j^{GBN} \cdot \vec{r} + \varphi). \tag{12}$$

The wave vector modulations $\vec{Q}_j^{GBN} = \frac{4\pi}{\sqrt{3}L_{GhBN}}\left[\cos\frac{2\pi(j-1)}{3} + \theta, \sin\frac{2\pi(j-1)}{3} + \theta\right]$ are not commensurate with the triangular lattice formed in GG. The potential is parametrized by the angle $\theta$, the amplitude $U_0$, the length scale $L_{hBN}$ and the phase $\theta$. The relaxed lattice structure is obtained by minimizing the total energy:

$$E_{tot}[\{\vec{u}\}] = \frac{K}{2}\sum_{<i,j>}(\vec{u}_i - \vec{u}_j)^2 + \sum_i U(\vec{R}_i + \vec{u}_i). \tag{13}$$

To characterize the crossover between the commensurate to the incommensurate regimes we measure the structure factor:

$$S(\vec{q}) = \left|\frac{1}{N}\sum_i^N e^{-i\vec{q}\cdot(\vec{R}_i+\vec{u}_i)}\right|^2. \tag{14}$$

We consider deviations from the 60-degree commensurate configurations (*10*) that is realized for $\theta_{GG} = 1.1$ and $\theta_{GBN} = 0.55$. The evolution of the lattice structure for different values of $\theta_{GBN}$ in the magic angle region is shown in the supplementary material. We set the ratio between the elastic energy and amplitude of the hBN potential to $U_0/(KL_{TBG}) = 0.1$. In addition to the simulated cases in Fig 4, simulation for additional twist angle combinations can be seen in Fig S11.

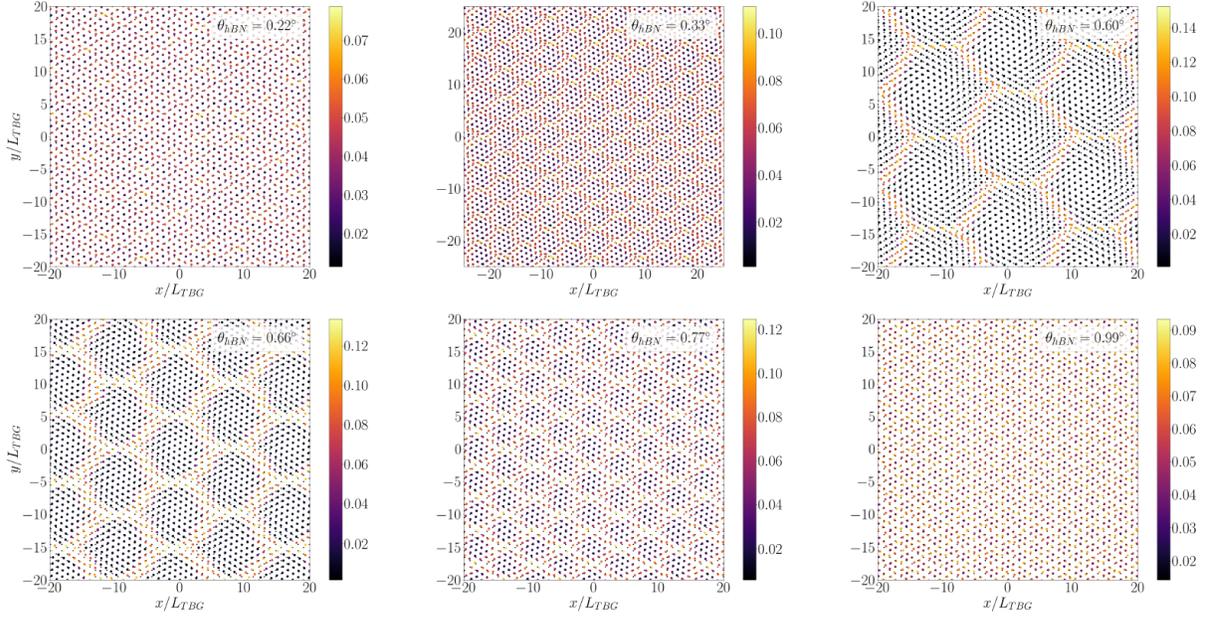

**Fig. S11. Additional 2D FK-model model simulations.** Fixing $\theta_{GG}$ =1.10º we change the relative angle between bottom graphene and the hBN $\theta_{GBN}$. The $\theta_{GBN}$ used for each simulation is marked on the upper right corner of every panels. The color scale represent the local displacement $|\delta u|/L_M$ of GG lattices after self-alignment.

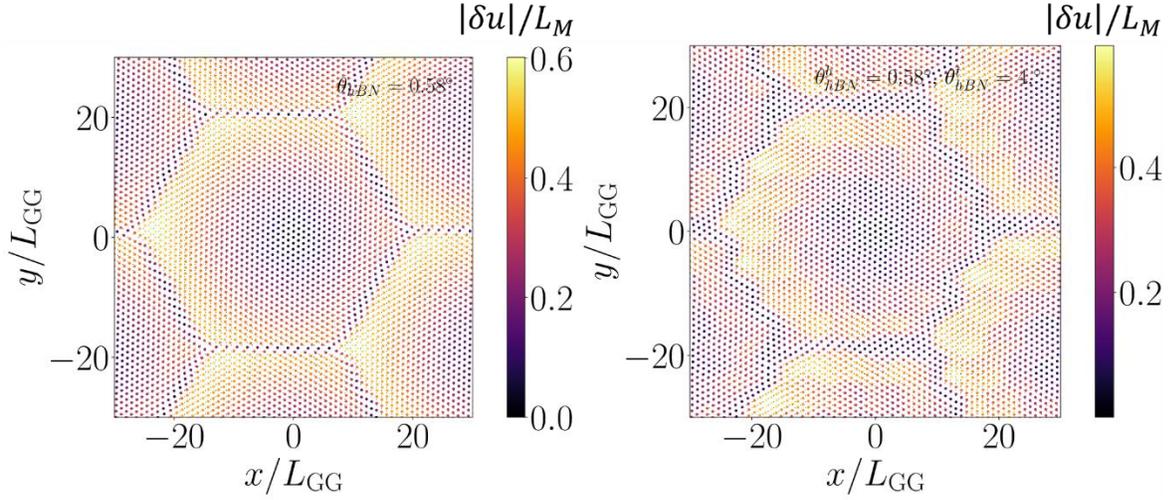

**Fig. S12. 2D FK-model model simulations comparison between TBG suspended (left) and encapsulated (right) on hBN encapsulated.** In the encapsulated geometry we fix $\theta_{GG}$ =1.10º, we assume aligned bottom hBN $\theta_{GBN,1} = 0.58$º while top hBN is rotated by an angle $\theta_{GBN,2} = 4$º away from the commensurate regime. Moreover, we set the amplitude of the bottom potential to $\frac{U_1}{L_{GG}Y} = 0.1$ and $\frac{U_2}{L_{GG}Y} = 0.03$ for the top one assuming the latter has a smaller adhesion energy. The color scale represent the local displacement of GG lattices after self-alignment $|\delta u|/L_M$.

The contribution of a top hBN capping layer the is misaligned to GG and bottom hBN can also be modeled. We find that capping hBN introduces additional disorder to the commensurate domains as presented in Fig S12.

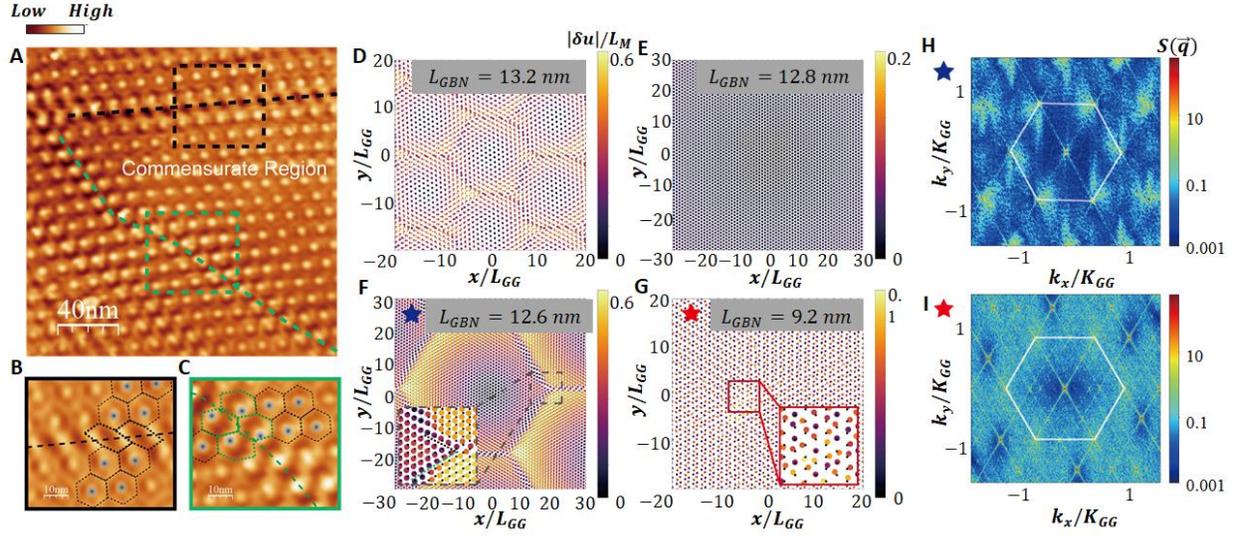

Fig. S13. **2D FK-model model simulation for quasicrystal before and after alignment.** (A) STM topography of commensurate double moiré patterns (L_M=12.1 nm) where 2 types of domain boundaries are observed between different commensurately stacked domains. (B) & (C) Zoom-in of the (B) "tensile" and (C) "shear" domain boundaries from the purple and green rectangle regions in A. The two types of domain boundaries are marked with purple and green dashed lines, respectively. GBN unit cells are outlined by the dashed hexagons and the superlattice distortion is clearly seen near the boundaries. The AA sites are marked by blue spots. The scale bars are 10nm. Tunneling conditions: VB = -500mV; VG=0; I = 20pA (D)-(G) Simulated moiré-of-moiré domains post relaxation based on FK-model for L_GG=12.8 and several values of L_GBN (D) L_GG≲L_GBN with L_GBN=13.2 nm, I L_GG=L_GBN " with " L_GBN=12.8 nm, (F) L_GG≳L_GBN with L_GBN=12.6 nm, (G) L_GG>L_GBN "with " L_GBN=9.8 nm. Color scales represent the local magnitude of moiré movement from their initial position before relaxation. The inset of (F) shows the two kinds of domain boundaries observed in the experiment: tensile and shear marked by black and green lines respectively. The inset of (G) reflects the local disorder due to the quasicrystal nature of the large incommensurability which resembles the observation in Fig 2I. (H) & (I) Structure factor $S(\vec{q})$ calculated for (F) & (G) respectively. 1st Brillouin zone of GBN is marked by the grey hexagon. In Fig. (I) additional non commensurate peaks originating from the interaction with BN are shown in blue while the GG periodicity in white.

13. **Discussion of the Phase Diagram**

The commensurate-incommensurate transition has been observed previously in GBN moiré pattern (*15*). Similarly, the FK-model is also employed to explain this transition when decreasing $\theta_{GBN}$ reduces the incommensurability and majority area of a GBN moiré unit cell relax to the preferred $C_B$ stacking. In GG-GBN system with small $\theta_{GBN}$ presented in this work, self-alignment constructs locally commensurate domains of the preferred AAB stacking near the $L_{GG}:L_{GBN} = 1:1$ commensuration, as shown in the phase diagram of Fig 3A in main text. Evidence shows the self-alignment behavior is universal in systems with coexistent moiré patterns with similar wavelengths and is also observed in twisted tri-layer graphene with relatively small twist angles ($\theta_{GG}$) (*16*). On the left of the 1:1 line of phase diagram in Fig 3A, $L_{GG}$ is small and $\theta_{GG}$ is large. The size of GG AA sites is significantly smaller than the GBN unit cells in the examples shown in Fig S5. In this regime the two incommensurate moiré patterns do not self-align possibly because the small wavelength lacks flexibility to align. In this case we observe two clear sets of Brag peaks for GG and GBN indicating that the observed patterns are spanned by 4 independent wavevectors. This regime is comprised mostly of of moiré quasicrystals (*17*).

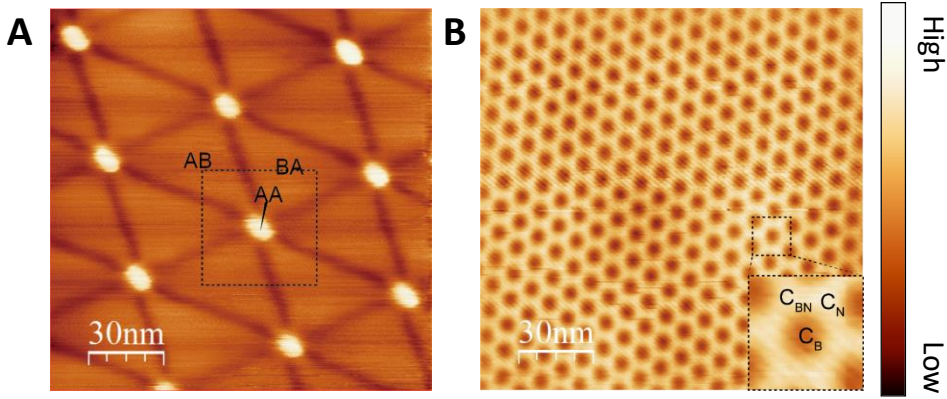

**Fig. S14. Topography of GG and GBN moiré patterns. (A)** Topography of a small angle GG with $\theta_{GG} = 0.28°$ not aligned to hBN. Different regions are marked as AA, AB and BA. **(B)** Topography of Monolayer Graphene-hBN moiré patterns as hexagons with $\theta_{GBN} = 1.42°$. Different regions are marked as $C_B$, $C_{BN}$ and $C_N$ in the inset. Topography scans are acquired with $V_G = 0$; $V_B = -300mV$; $I = 20pA$

However, on the right side of the 1:1 line where $\theta_{GG}$ is small and $L_{GG}$ is large, self-alignment becomes favorable as the size of GG AA sites saturate to about the same as that of the GBN $C_B$ sites due to the expansion of GG AB/BA sites, as Fig S13A demonstrates. Local AAB stacking is achieved for almost all GG AA sites in the topographies shown within Fig S3. In addition we find commensurate regions with $L_{GG}:L_{GBN} = 2:1, 3:1 \ldots$ Examples are shown in Fig S4 and S6. In addition, the alignment between incommensurate GG and GBN sites introduces a larger unit cell whose wavelength is an integer multiple of the smaller wavelength ($L_{GBN}$) such that local translational symmetry is broken. This broadens the GG peaks in the FFTs of Fig S6A to S6E. Additional features in FFTs of Fig S6F to S6L arising at periods between the GG and GBN suggest that the moiré-moiré interaction modifies the detailed structure of the composite moiré patterns. The interplay of self-alignment and the magnified strain effect in this regime may introduce disordered phases such as Bragg glass or geometrically frustrated quasicrystals (*18, 19*).

**Supplementary References:**